\documentclass[11pt,a4paper]{article}
\usepackage{jinstpub}
\arxivnumber{2205.04549}

\usepackage{booktabs}
\usepackage{array}
\usepackage{bm} 
\usepackage{graphicx} 
\usepackage{dcolumn} 

\usepackage[english]{babel}
\usepackage[T1]{fontenc} 
\usepackage[utf8]{inputenc} 

\usepackage{adjustbox} 
\usepackage{tabularx}
\usepackage{makecell} 
\usepackage[xcolor={table, dvipsnames}, ulem]{changes}
\definechangesauthor[color=orange]{ML}
\definechangesauthor[color=blue]{IN}
\usepackage{derivative}
\usepackage{lineno}
\usepackage[separate-uncertainty=true]{siunitx}[=v2]
\usepackage[caption=false]{subfig}
\newcolumntype{L}{>{$}l<{$}} 
\newcolumntype{C}{>{$}c<{$}} 
\newcolumntype{R}{>{$}r<{$}} 

\begin{document}

\newcommand{\nvbb}{\ensuremath{0 \nu \beta \beta}}
\newcommand{\Vb}{\ensuremath{V_{\rm{bol}}}}
\newcommand{\Rb}{\ensuremath{R_{\rm{bol}}}}
\newcommand{\vbol}{\ensuremath{v_{\rm{bol}}}}
\newcommand{\rbol}{\ensuremath{r_{\rm{bol}}}}
\newcommand{\RL}{\ensuremath{R_{\rm{bias}}}}

\newcommand{\eqnref}[1]{eq.~(\ref{#1})}
\newcommand{\eqnsref}[1]{eqs.~(\ref{#1})}
\newcommand{\secref}[1]{Section~\ref{#1}}
\newcommand{\tabref}[1]{Table~\ref{#1}}
\newcommand{\figref}[1]{Figure~\ref{#1}}

\def\verticaldistance{10pt}

\title{An Energy-dependent Electro-thermal Response Model of CUORE Cryogenic Calorimeter} 

\collaboration{CUORE collaboration}
\collaboration{CUORE Collaboration}
\author[a]{D.~Q.~Adams,}
\author[a]{C.~Alduino,}
\author[b]{K.~Alfonso,}
\author[a]{F.~T.~Avignone~III,}
\author[c]{O.~Azzolini,}
\author[d]{G.~Bari,}
\author[e,f]{F.~Bellini,}
\author[g]{G.~Benato,}
\author[h]{M.~Beretta,}
\author[i]{M.~Biassoni,}
% ...

\emailAdd{cuore-spokesperson@lngs.infn.it}

\affiliation[a]{Department of Physics and Astronomy, University of South Carolina, Columbia, SC 29208, USA}
\affiliation[b]{Department of Physics and Astronomy, University of California, Los Angeles, CA 90095, USA}
\affiliation[c]{INFN -- Laboratori Nazionali di Legnaro, Legnaro (Padova) I-35020, Italy}
\affiliation[d]{INFN -- Sezione di Bologna, Bologna I-40127, Italy}
\affiliation[e]{Dipartimento di Fisica, Sapienza Universit\`{a} di Roma, Roma I-00185, Italy}
\affiliation[f]{INFN -- Sezione di Roma, Roma I-00185, Italy}
\affiliation[g]{INFN -- Laboratori Nazionali del Gran Sasso, Assergi (L'Aquila) I-67100, Italy}
\affiliation[h]{Department of Physics, University of California, Berkeley, CA 94720, USA}
\affiliation[i]{INFN -- Sezione di Milano Bicocca, Milano I-20126, Italy}
% ...

\date{\today}

\abstract{
The Cryogenic Underground Observatory for Rare Events (CUORE) is the most
sensitive experiment searching for neutrinoless double-beta decay
($\nvbb$) in $^{130}\text{Te}$. CUORE uses a cryogenic array
of 988 TeO$_2$ calorimeters operated at
$\sim$10 mK with a total mass of 741 kg. To further increase the
sensitivity, the detector response must be well understood. Here, we present a
non-linear thermal model for the CUORE experiment on a detector-by-detector
basis. We have examined both equilibrium and dynamic electro-thermal models of
detectors by numerically fitting non-linear differential equations to the
detector data of a subset of CUORE channels which are well
characterized and representative of all channels.
We demonstrate that the hot-electron effect
and electric-field dependence of resistance in NTD-Ge thermistors alone are
inadequate to describe our detectors' energy dependent pulse shapes. We
introduce an empirical second-order correction factor in the exponential
temperature dependence of the thermistor, which produces excellent agreement
with energy-dependent pulse shape data up to $\SI{6}{\mega \electronvolt}$.
We also present a noise analysis using the fitted thermal
parameters and show that the intrinsic thermal noise is negligible compared to
the observed noise for our detectors.
}

\maketitle 
\flushbottom


\section{Introduction}\label{sec:intro}

Neutrino-less double beta decay ($\nvbb$) is a second-order nuclear process
which, if observed, will violate the lepton number conservation in the Standard
Model of particle physics. An observation of $\nvbb$ decay would imply that at
least one neutrino is Majorana in nature \cite{schechter1982}, place
constraints on the neutrino masses \cite{dolinski2019}, and will have
fundamental implications for both neutrino and beyond Standard Model physics
\cite{moe1994}. A constraint on the $\nvbb$ decay half-life, even if the
process is not directly observed, can still be converted into an inference on
the effective neutrino mass\cite{vergados2016}. The process also provides
insight on how the imbalance between matter and anti-matter was created in
early universe \cite{fukugita1986}.

CUORE (Cryogenic Underground Observatory for Rare Events) is an ongoing
tonne-scale experiment \cite{martinez2017} searching for $\nvbb$ in $^{130}$Te.
The detector contains a total 19 towers of $\text{TeO}_{2}$ crystals, each
tower consisting of 52 $5 \times 5 \times 5 \ \SI{}{\centi \meter^3}$ cubic
crystals. The low heat capacity of the crystals at cryogenic temperatures
$\sim$10~mK provides measurable temperature changes for energy deposition
events. The $\text{TeO}_{2}$ crystals are used as the source of double beta
decays \cite{alfonso2015} coming from $^{130}\text{Te}$.

CUORE has placed a 90\% Confidence Interval (CI) lower limit on the half life
of $^{130}\text{Te}$ $\nvbb$ as $\SI{2.2e25}{yr}$, a world-leading one
regarding the process in $^{130}\text{Te}$ \cite{thecuorecollaboration2022}.
CUORE's detection principle, the calorimetric approach, takes advantage of the
fact that the energy for each detected radioactive event causes a sudden
temperature increase in the $\text{TeO}_{2}$ crystal (absorber). CUORE uses
Neutron Transmutation Doped (NTD) Germanium thermistors glued to the crystal to
detect the subtle change in temperature \cite{haller1984ntd,wang1990}.

If the energy deposited is small enough, the response of the detector system
can be solved exactly using a system of linear differential equations, with the
signal pulse rise and decay time constants being a combination of different
eigenvalues of the coefficient matrix. However, in CUORE there have been
observed changes in pulse shape for events across a wide range of energies
\cite{carrettoni2011}, from few tens of keV to $\SI[]{6}[]{\mega
    \electronvolt}$. We ascribe such changes to a different response of the whole
system beyond the small-signal limit for high energy depositions. Since the
region of interest (ROI) for CUORE is around\cite{thecuorecollaboration2022}
$Q_{\beta \beta} = \SI{2527.518\pm0.013}{\kilo \electronvolt}$, which is in the
middle of our energy range, it is imperative for us to better understand the
behavior of the detector. Thus, we have developed an energy-dependent thermal
model with physical parameters. Furthermore, because the energy interpreted for
a generic calorimetric detectors depends on the pulse height, we find it
advantageous to build a generic mathematical model framework for these
calorimetric detectors. For CUORE, understanding the detector response is a
prerequisite for pulse shape analysis (PSA). PSA enables effective filtering of
energy-dependent pulse shapes, discriminates energy deposition based on
particle type \cite{gironi2012} and allows efficient triggering at low
energies. With a model for the detector response, we can reproduce the pulses
with higher fidelity and lower noise, potentially reducing timing jitter for
each event. In addition, a model that describes the pulse shapes will be
beneficial for machine learning algorithms for training and classification
purposes.

The non-linear model is built on previous efforts that studied NTD-Ge
thermistors in the linear (small-signal) regime. \cite{mccammon2005, wang1990,
soudee1998, zhang1998, piat2001}, and is able to predict the energy-dependence
of the pulse shape. We adopt a combined model of both electrical field effect
and hot electron effect to describe the electro-thermal property of the NTD-Ge
semiconductor. In the thermal circuit, we also account for
temperature-dependent heat capacities and thermal conductance. The simulated
response is shown to be in close agreement with the observed signals across an
extensive energy range if we include an additional second-order temperature
correction to the NTD-Ge resistivity.

We conclude with a discussion on the noise response of the system based on the
fitted model parameters. We observe excess noise in the CUORE measured noise
power spectrum compared to the simulation. In our simulated noise model, a
dominant contributor is the biasing resistor in the electrical circuit.
Assuming additional $1/f$ and linear noise, the noise model can reproduce the
continuous noise power spectra. However, because the CUORE electronics was
designed to mitigate the $1/f$ noise, we are investigating the sources of the
additional noise terms.

\section{Electro-thermal Models}\label{sec:tm_models}

Thermal modeling for macrocalorimeters applies classical thermodynamics to the
components of the calorimeters: the readout device (e.g. a thermistor) and its
electrical circuit. Previous studies have extensively covered near-equilibrium
calorimeters
\cite{galeazzi2003,mccammon2005,figueroa-feliciano2006,Mather:84,mather1982,mather1984,Jones:53}.
These studies provide comprehensive techniques of analyzing sensitive
thermistors in small excursion from their equilibrium states, resulting in a
handful of useful linear theories. However, the linear thermal models have a
limitation that the response pulse shape is independent of event energy. We
wish to find a non-linear model so that the pair annihilation peak,
fully-contained high-energy $\gamma$ events (around $1$ to $\SI{2}{\mega
    \electronvolt}$) and $\alpha$ signals (around $\SI{6}{\mega \electronvolt}$)
can be reproduced with the same set of parameters, similar to one of the
previous studies \cite{alessandrello1993} but in finer detail. We go to second
order Taylor expansion near the equilibrium point of the detector to study the
energy-dependent response of the calorimeter.

\subsection{Electrical and Thermal Circuit}\label{sec:circuits}
\begin{figure}
    \centering
    \includegraphics[width = 0.35\textwidth]{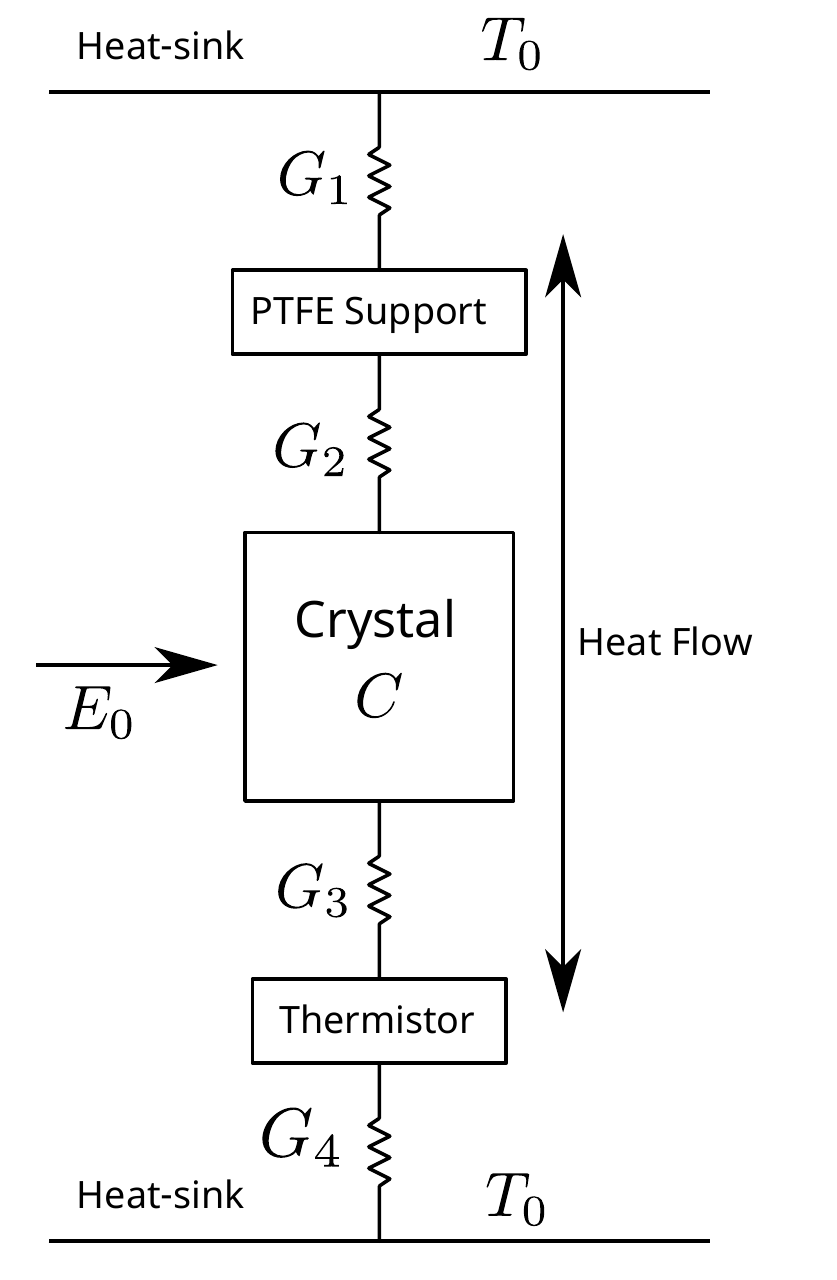}
    \caption{Simplified block diagram for CUORE thermal model. The crystal is modeled as a single object with heat capacity $C$ coupled to the heat-sink (with constant temperature $T_0$) through the PTFE support. When there is an energy deposition $E_0$ the heat flows through the PTFE support ($G_1$, $G_2$) and/or the NTD gold wires ($G_4$) and eventually to the heat-sink.}
    \label{fig:block_diagram}
\end{figure}

Each CUORE readout channel consists of a $\text{TeO}_{2}$ crystal, an NTD-Ge
thermistor for temperature readout, a silicon-based heater for thermal gain
stabilization, and several PTFE spacers for isolating the crystal from the
heat-sink. The NTD-Ge thermistors have a dimension of $3.0 \times 2.9 \times
    0.9 \ \SI{}{\milli \meter^3}$ ($L \times W \times H$) \cite{Alduino_2016}.
\figref{fig:block_diagram} provides a qualitative description of the energy
deposition process: the temperature of the crystal rises and so does the
temperature of the thermistor system. The small heat capacity of the thermistor
ensures that it closely follows the crystal temperature. In a few seconds,
deposited energy flows out of the system through the thermal paths from the
detector to the heat sink (the PTFE support and/or the NTD gold wires).

\begin{figure}
    \centering
    \includegraphics[width = 0.45\textwidth]{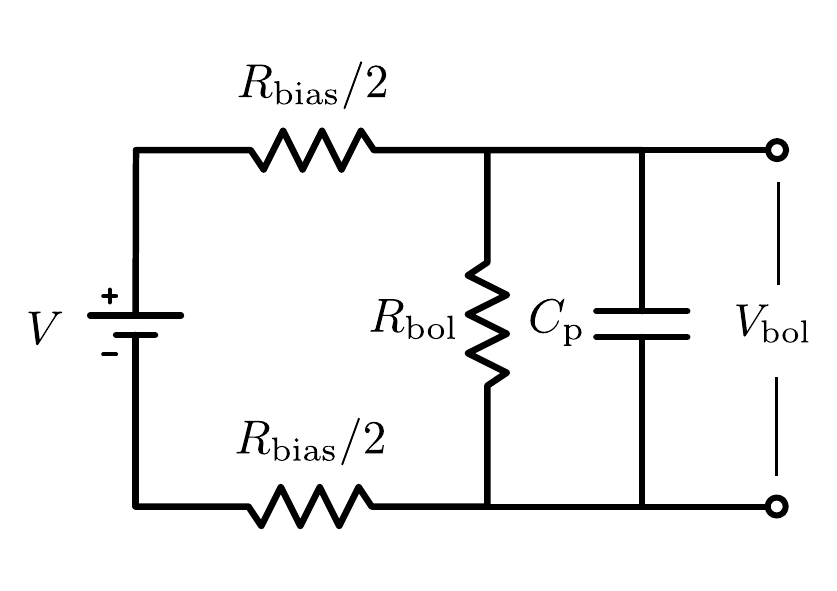}
    \caption{CUORE Detector electrical circuit. The two bias resistors before and after the NTD-Ge, labelled $\Rb$, can be combined into one equivalent resistor $R_{\rm{bias}}$.}
    \label{fig:electrical_circuit}
\end{figure}

We start from a common readout circuit for thermal detectors as shown in
\figref{fig:electrical_circuit}, where the bias resistor, together with a
biasing voltage, determines the operational resistance of the detector. The
bias resistor should be large compared to the thermistor so that the biasing
circuit acts as a quasi-constant current source. In the case of extra energy
deposition, the thermistor's resistance decreases as its temperature increases.
In this approximation, the current through the thermistor is held constant, and
the thermistor's self-heating power decreases along with the decrease in
resistance, thus forming a negative thermal feedback that tends to stabilize
the system. Realisitically, in such systems, the parasitic capacitance is also
seen as a load, inserting possible instabilities at the frequency where the
modulus of its impedance is equal to the thermistor impedance.

Additionally, the readout circuit, positioned outside the cryostat at 300K,
holds some residual cable capacitance. The parasitic capacitance was measured
and determined \cite{Andreotti_2009, Arnaboldi_2018} to be $C_{\textrm{p}}
    \approx \SI{500}{\pico \farad}$ between the detector cold stages at
$\SI{10}{\milli \kelvin}$ and the front-end (FE) board. Therefore, we add the
wire capacitance here in parallel with the thermistor in
\figref{fig:electrical_circuit}. The corresponding electrical equation is:
\begin{equation}\label{eqn:electrical_1}
    C_{\textrm{p}} \RL \dot{V}_{\rm{bol}}(t) +\frac{\Vb(t)}{\Rb(t)} \RL + \Vb(t) - V = 0,
\end{equation}
where $V$ is the constant voltage source bias.
We note that the thermistor $\Rb(\Vb(t), T(t))$ is temperature and
voltage dependent.
This is our window to the thermal circuit.

Additionally, the output signal (NTD-Ge voltage $\Vb$) is amplified and
filtered through a 6-order Bessel filter. The transfer function of the Bessel
filter is \cite{Arnaboldi_2018,vignati2009}:
\begin{equation}
    H(\sigma) = \frac{10395}{\sigma ^{6} +21\sigma ^{5} +210\sigma ^{4} +1260\sigma ^{3} +4725\sigma ^{2} +10395\sigma +10395},
\end{equation}
with $\sigma = 2.703395061j\omega / (2\pi f_{c})$ and $f_{c} = \SI{120}{Hz}$.
All our simulations producing $\Vb$ are digitally amplified and filtered to
match the true amplification and Bessel filter in the electronics readout
chain.

The thermal power between materials of different temperatures usually takes the
form of:
\begin{equation}\label{eqn:th_cond}
    P_{i-j} = \int_{T_j}^{T_i} dT \, G_{i-j}(T) = g_{i-j} ( T_i^{a_{i-j}} -  T_{j}^{a_{i-j}}),
\end{equation}
with the power flowing from node $i$ to node
$j$ in the above equation. We use
$G$ to denote the thermal conductance at
a certain temperature, and $g$ to represent the thermal
conductivity coefficient.
\begin{figure*}
    \centering
    \includegraphics[width = 0.75\textwidth]{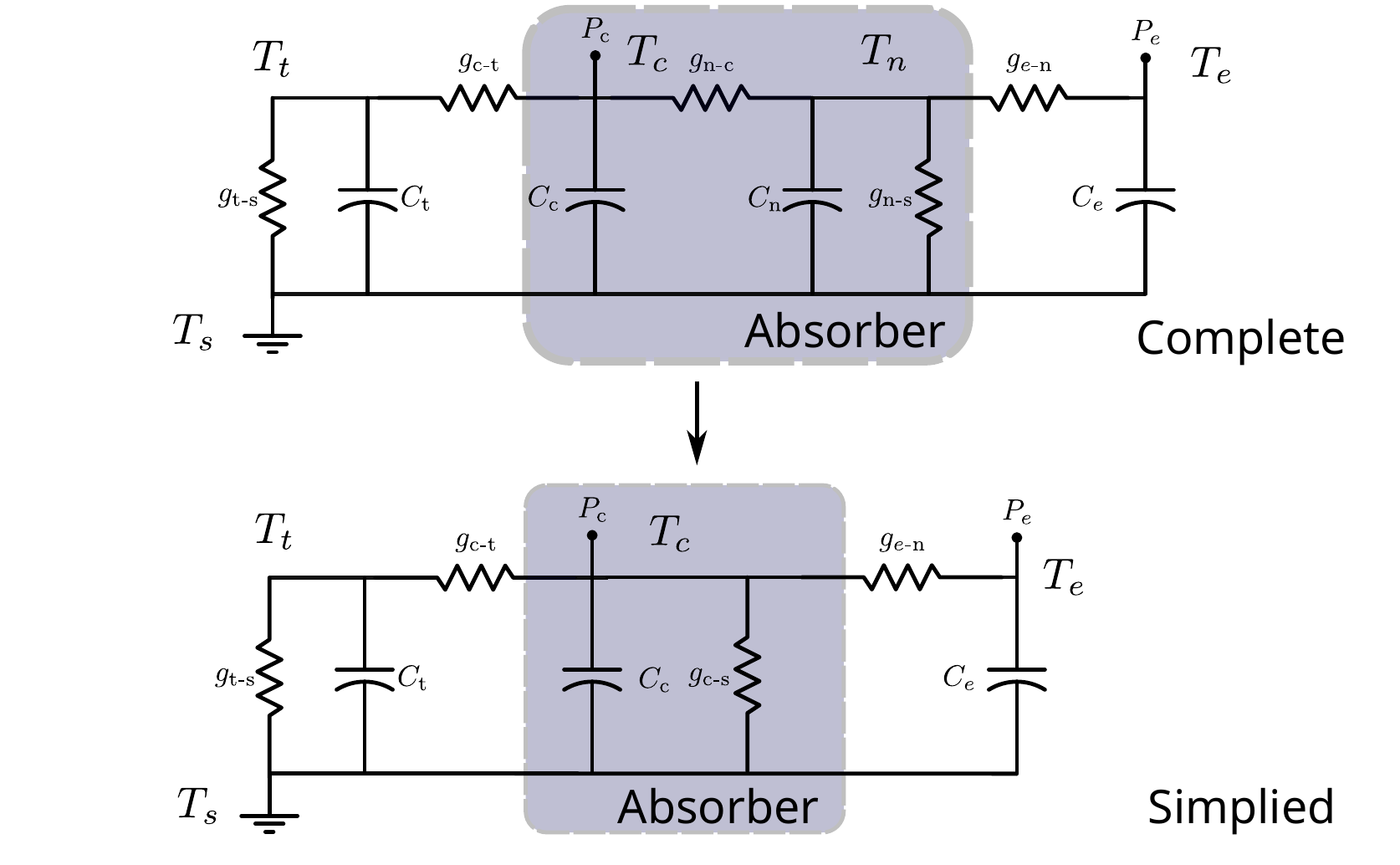}
    \caption{Thermal circuit for a single CUORE unit. The top diagram shows the components: PTFE support, the $\text{TeO}_{2}$ crystal, the NTD-Ge lattice, and the Ge electrons. The bottom diagram shows the crystal node simplified with the combination of the $\text{TeO}_{2}$ and NTD-Ge lattice.}
    \label{fig:thermal_circuit}
\end{figure*}

\figref{fig:thermal_circuit} shows the complete thermal circuit (top) and
its
simplified version (bottom) used in this work.
The illustrated nodes from left to right are: PTFE support for the crystal, the
$\text{TeO}_{2}$ crystal absorber, the NTD-Ge lattice glued to the
crystal surface, and the electron
system in the NTD-Ge lattice.
There are gold wires connecting the NTD-Ge lattice to
the copper frame (heat-sink) for NTD-Ge biasing and read-out, and thus we
represent the conductivity term
by its coefficient $g_{n-s}$.
Similarly, $g_{n-c}$ represents the glue conductivity term
connecting
the NTD-Ge and the $\text{TeO}_{2}$ crystal, and
$g_{e-n}$ represents
the electron-phonon thermal coupling coefficient within the NTD-Ge sensor.
We have neglected the thermal coupling from the Ge electrons directly to the
heat-sink,
since its effect is likely to be small compared to other thermal couplings,
mainly the ones denoted by $g_{n-c}$,
$g_{n-s}$, and $g_{e-n}$
\cite{pedretti2004,vignati2009}.
We further simplify the model by absorbing the NTD-Ge lattice node into the
$\text{TeO}_{2}$ crystal node, as shown in the bottom part of
\figref{fig:thermal_circuit},
effectively assuming that both nodes will always be at similar temperatures.
We justify this by noting the tiny heat capacity of the NTD-Ge lattice,
compared to that of the $\text{TeO}_{2}$ crystal, and the large
thermal coupling between the Ge and the
$\text{TeO}_{2}$ due to the glue.
We have tabulated the estimated heat capacities for the calorimetric components
in \tabref{tab:component_hea_cap}.
Similarly, the effect of
glue conductance is absorded in $g_{e-n}$, which now
represents an effective conductance between the sensor electron system and the
main absorber crystal.
\begin{table}
    \caption{Table of estimated heat capacities of different components for CUORE crystals and NTD-Ges. The tiny capacity of the NTD-Ge lattice compared to the crystal and the electron system justifies its assimilation into the crystal node. The temperature dependence of the heat capacities are also revealed in the capacity matrix $C(T + \delta y)$. }\label{tab:component_hea_cap}
    \centering
    \adjustbox{width=0.50\textwidth}{
    \begin{tabular}{cc}
        \noalign{\vskip 1mm}
        \hline
        \hline
        Component                                    & \makecell{Estimated Heat Capacity at    \\ $T = \SI{10}{\milli \kelvin}$ $[\si{\joule / \kelvin}]$} \\[2ex]
        \hline
        \noalign{\vskip 1mm}
        Electrons \cite{olivieri2006,aubourg1993}    & $\num{1e-10} \propto T$\footnotemark[1] \\
        NTD-Ge Lattice \cite{keesom1959,ventura2008} & $\num{2.5e-14} \propto T^3$             \\
        Crystal \cite{barucci2001,ventura2008}       & $\num{2.2e-9} \propto T^3$              \\
        PTFE Support \cite{singh2015,Alduino_2016}   & $\num{2.1e-8} \propto T$                \\
        \hline
        \hline
    \end{tabular}
    }
    \footnotetext[1]{We acknowledge the possibility
    of Schottky anomaly\cite{ventura2008} $C_e \sim \Gamma T_e + A / T_e^2 $
    but in our fits we were not able to determine parameter $A$.}
\end{table}

The combined node will have an additional thermal coupling to the heat-sink due
to the gold wires bond to the Ge lattice. For book-keeping purposes, this
conductivity coefficient is relabelled as $g_{c-s}$, and the electron-phonon
coupling in the NTD-Ge lattice is relabelled as $g_{e-c}$. Apart from keeping
the number of free parameters to a minimum, there is another reason for using
three thermal nodes: in small signal limit, where the linear theory and
solutions can be applied, the solution in the frequency domain with four poles
and one zero is able to describe the CUORE pulses
\cite{nutini2018,alfonso2020}. Our choice of three thermal nodes plus one
electrical read-out node fits comfortably in this picture.

\subsection{Circuit Equations}
We label the self-heating of the thermistor as $P_{e}$ and the power deposited
on the crystal as $P_{c}$. Reading from the circuit, we have the following set
of equations:
\begin{equation}\label{eqn:thermal_1}
    C_{e}\dot{T}_{e} = P_{e} - g_{e-c}\left( T^{a_{e-c}}_{e} -T^{a_{e-c}}_{c}\right),
\end{equation}
\begin{eqnarray}\label{eqn:thermal_2}
    C_{c}\dot{T}_{c} = P_{c} && + g_{e-c}\left( T^{a_{e-c}}_{e}  -T^{a_{e-c}}_{c}\right) \nonumber \\
    && - g_{c-t}\left( T^{a_{c-t}}_{c} -T^{a_{c-t}}_{t}\right) \nonumber \\
    && - g_{c-s}\left( T^{a_{c-s}}_{c} -T^{a_{c-s}}_{s}\right),
\end{eqnarray}
\begin{eqnarray}\label{eqn:thermal_3}
    C_{t}\dot{T}_{t} = && g_{c-t}\left( T^{a_{c-t}}_{c} -T^{a_{c-t}}_{t}\right) \nonumber \\
    && - g_{t-s}\left( T^{a_{t-s}}_{t} -T^{a_{t-s}}_{s}\right).
\end{eqnarray}

If we expand \eqnref{eqn:electrical_1} near the equilibrium state (noted as
"eq"), and denote:
\begin{eqnarray*}
    \Rb && \to \overline{R}_{\rm{bol}} (eq) + \delta \rbol (t), \\
    \Vb && \to \overline{V}_{\rm{bol}} (eq) + \delta \vbol (t),
\end{eqnarray*}
we obtain a dynamic equation describing the changes in the terms of
interest $\Rb$ and $\Vb$:
\begin{equation}\label{eqn:electrical_2}
    C_{\textrm{p}} \RL \dot{\delta v}_{\rm{bol}} +\frac{\delta \vbol \overline{R}_{\rm{bol}} -\overline{V}_{\rm{bol}} \delta \rbol}{\left(\overline{R}_{\rm{bol}} + \delta \rbol \right)\overline{R}_{\rm{bol}}} \RL + \delta \vbol = 0.
\end{equation}
We will present the thermistor's dependence of $T_{e}$ and
$\Vb$
later in \secref{sec:ntd}.

Regarding the thermal circuits \eqnsref{eqn:thermal_1} to
(\ref{eqn:thermal_3}), we follow a similar expansion around equilibrium to
arrive at:
\begin{equation}
    C \dot{\delta y} = \Delta P + \delta P_{i-j},
\end{equation}
with
\begin{eqnarray}
    \delta P_{i-j} = && g_{i-j}\left[ (\overline{T}_{i} + \delta t_i)^{a_{i-j}} - (\overline{T}_{j} + \delta t_j)^{a_{i-j}} \right] \nonumber \\
    && - g_{i-j}\left[ \overline{T}_{i}^{a_{i-j}} - \overline{T}_{j}^{a_{i-j}} \right],
\end{eqnarray}
where $\delta t_{i}$ and $\delta t_{j}$ are the small
changes in temperature from equilibrium $\overline{T}_{i}$,
$\overline{T}_{j}$, and $\delta y$
represents a general vector of excursions from the equilibrium. The heat
capacities
$C$ are
functions of
temperature: $C=C(\overline{T}(eq) + \delta y)$.
The $\Delta P$ term specifically denotes the electro-thermal
feedback from
the Joule heating of the NTD-Ge thermistor. This is the change in heating power
of the thermistor when there is a change in electron temperature or voltage
across the thermistor. Quantitatively:
\begin{eqnarray}\label{eqn:delta_p}
    \Delta P && = \frac{( \overline{V}_{\rm{bol}} +(\delta \vbol))^{2}}{\overline{R}_{\rm{bol}} + (\delta \rbol)} - \frac{\overline{V}_{\rm{bol}}^{2}}{\overline{R}_{\rm{bol}}} \nonumber \\
    && = \frac{\overline{V}_{\rm{bol}}^2  + 2 \overline{V}_{\rm{bol}} (\delta \vbol) + (\delta \vbol)^2}{\overline{R}_{\rm{bol}} + (\delta \rbol)} - \overline{P}.
\end{eqnarray}
In general we find a system of equations in the form:
\begin{equation}\label{eqn:gen_sys}
    C(\overline{T} + \delta y) \dot{\delta y} = f(\delta y) + x,
\end{equation}
with
\begin{equation}
    \delta y=\begin{bmatrix}
        \delta \vbol \\
        \delta t_{e} \\
        \delta t_{c} \\
        \delta t_{t}
    \end{bmatrix}, C =\begin{bmatrix}
        C_{\rm{p}}                           \\
        C_{e}(\overline{T_e} + \delta t_{e}) \\
        C_{c}(\overline{T_c} + \delta t_{c}) \\
        C_{t}(\overline{T_t} + \delta t_{t})
    \end{bmatrix}.
\end{equation}
In \eqnref{eqn:gen_sys}, $x$ is some additional
injected power vector,
such as the power injection due to some energy deposition in the
crystal. The full form will be
presented at the end of \secref{sec:ntd}.

\subsection{NTD-Ge Characteristics}\label{sec:ntd}
NTD-Ge resistivity depends on its charge-carrier temperature ($T_{e} =
    \overline{T}_{e} + \delta t_{e}$) and applied voltage ($V_{\rm{bol}} =
    \overline{V}_{\rm{bol}} + \delta \vbol$) across the chip. Since the temperature
dependence of NTD-Ge is exponential, the behavior of the thermistor is expected
to make a large contribution to the non-linearity we aim to study.

Starting from the Shklovskii-Efros law \cite{mccammon2005,Efros_1975} we obtain
thermistor resistivity under zero bias voltage:
\begin{equation}\label{eqn:ohmic_vrh}
    R(T_{e} )= R_{0}\exp\left(\frac{T_{0}}{T_{e}}\right)^{\gamma }.
\end{equation}
For low temperatures, $\gamma = 0.5$ is the common value in
literature \cite{mccammon2005a}.
It has also been observed that for extremely
low temperatures (such as 10 mK), even injected power as low as
\(10^{-14} \ \SI{}{\watt}\)
can induce non-ohmic behavior (i.e. deviation from
\eqnref{eqn:ohmic_vrh})
on the thermistor's voltage-current (V-I) curve
\cite{ventura2008}.
Usually such deviation at low temperature is attributed to electron-phonon
thermal decoupling
between the electrons and the NTD-Ge lattice \cite{wang1990} when
the electrical field is small. That is, there exists a
thermal resistance between the electrons and the NTD-Ge lattice, and
their temperatures are related by:
\begin{equation}\label{eqn:hem}
    P_{e} = I_{\rm{bol}} V_{\rm{bol}} = g_{e-n} ( T_e^{a} -  T_{n}^{a} ) = g_{e-c}( T_e^{a} -  T_{c}^{a} ).
\end{equation}
We use the last equality when we combine the NTD-Ge lattice into the
$\text{TeO}_{2}$ crystal node.
Typical value for
$a$ is around 5 to 6 \cite{soudee1998}.
However, studies have found that considering
an additional term with
E-field-induced hopping conduction often gives better agreement with data
\cite{zhang1998,piat2001}.
We take the most widely accepted form derived by Hill
\cite{piat2001,hill1971}
for the modified
resistivity: \footnote{We note here that in original Hill paper the weak-field induced
hopping conduction is characterized by a $\sinh(x)$ function where $x = -eE L_{h} / k_{B} T_{e}$.
When converted to resistivity we should have $\rho = 1 / \sigma = E / J \sim \rho(x = 0) \cdot x / \sinh(x)$
behavior. In contrast experiments claim the modification on resistivity in the $x \gg 1$ region
to be $\rho \sim \rho(x = 0) \cdot 1 / \exp(x)$
which drops the non-exponential $x$ factor. We tested both models and found that
the exponential correction factor agrees better with our data.}
\begin{eqnarray}\label{eqn:electrical_vrh}
    R(\Vb ,T_{e}) && = R(0,T_{e}) \exp \left( \frac{-eE L_{h}}{k_{B} T_{e}} \right) \nonumber \\
    && = R_{0}\exp\left(\frac{T_{0}}{T_{e}}\right)^{0.5 }\exp\left(\frac{-e \Vb \lambda _{0}}{k_{B} T^{1.5}_{e} W}\right).
\end{eqnarray}
Here $L_h$ is the characteristic hopping length at a given
temperature,
which should,
in theory scale with \cite{kenny1989} $T^{- \gamma} = T^{-0.5}$.
After changing the field strength $E$ into
$\Vb / W$ ($W$ being the effective
thermistor width), we now describe the field correction term with a single
unknown parameter $\lambda_0$.

Expanding \eqnref{eqn:electrical_vrh} near equilibrium but keeping the
exponential, we obtain:
\begin{eqnarray}\label{eqn:resistance_expansion}
    \overline{R}_{\rm{bol}} + \delta \rbol && = R_{0}\exp\left(\frac{T_{0}}{\overline{T}_{e} +\delta t_{e}}\right)^{\gamma }\exp\left[\frac{-C_{\lambda }\left(\overline{V}_{\rm{bol}} +\delta \vbol \right)}{\left(\overline{T}_{e} +\delta t_{e}\right)^{\gamma +1}}\right]                                                                                                                                                                    \\
    && = R_{0}\exp\left[\left(\frac{T_{0}}{\overline{T}_{e}}\right)^{\gamma }\left( 1-\gamma \frac{\delta t_{e}}{\overline{T}_{e}} + ... \right)\right] \\ \nonumber
    &&  \qquad \qquad \times \exp\left[ -C_{\lambda }\frac{\overline{V}_{\rm{bol}}}{\overline{T}_{e}^{\gamma +1}} -C_{\lambda }\frac{\delta \vbol}{\overline{T}_{e}^{\gamma +1}} +C_{\lambda } (\gamma +1)\frac{\delta t_{e}\overline{V}_{\rm{bol}}}{\overline{T}_{e}^{\gamma +2}} + ... \right] \\
    && =\overline{R}_{\rm{bol}}\exp\left[ -\eta \left(\frac{\delta t_{e}}{\overline{T}_{e}} \right) + C_{\lambda } (\gamma +1)\frac{\overline{V}_{\rm{bol}}}{\overline{T}_{e}^{\gamma +1}} \left(\frac{\delta t_{e}}{\overline{T}_{e}}\right) \right. \\ \nonumber
    && \left. \qquad \qquad \qquad - C_{\lambda }\frac{\overline{V}_{\rm{bol}}}{\overline{T}_{e}^{\gamma +1}} \left(\frac{\delta \vbol}{\overline{V}_{\rm{bol}}} \right) + \mathcal{O}\left( \delta t_{e}^2, \delta \vbol^2, \delta t_{e} \delta \vbol \right) \right] \\
    && \equiv \overline{R}_{\rm{bol}}\exp(- \alpha), \label{eqn:resistance_expansion_sub}
\end{eqnarray}
where $\eta = \gamma \left(\frac{T_{0}}{T_{e}}\right)^{\gamma }$ is the
temperature sensitivity, and $C_{\lambda} = e \lambda_0 / k_{B} W$.
Substituting the above results into \eqnsref{eqn:electrical_2} and
(\ref{eqn:delta_p}), we have:
\begin{eqnarray}\label{eqn:electrical_3}
    C_{\textrm{p}} \RL \dot{\delta v}_{\rm{bol}} +\delta \vbol \frac{\RL}{\overline{R}_{\rm{bol}}} e^{\alpha } && -\overline{V}_{\rm{bol}}\frac{\RL}{\overline{R}_{\rm{bol}}}\left( 1-e^{\alpha }\right) \nonumber + \delta \vbol = 0.
\end{eqnarray}
The thermal equation regarding the electrons becomes:
\begin{equation}
    C_{e}\dot{\delta t}_{e} = \overline{P}_{e} \left( e^{\alpha } -1\right) + 2\delta \vbol \frac{\overline{V}_{\rm{bol}}}{\overline{R}_{\rm{bol}}} e^{\alpha } + \frac{\delta \vbol^2}{\overline{R}_{\rm{bol}}} e^{\alpha } - \delta P_{e-c}.
\end{equation}

We now summarize the system below (the bars are dropped for clarity):
\begin{equation}\label{eqn:ode_sys}
    C(T + \delta y) \dot{\delta y} \ = \begin{bmatrix}
        {\displaystyle -\delta \vbol\left(\frac{e^{\alpha }}{\Rb} +\frac{1}{\RL}\right) -\frac{\Vb}{\Rb}\left( e^{\alpha } -1\right)}                           \\[\verticaldistance]
        {\displaystyle P_{e}\left( e^{\alpha } -1\right) +\frac{2\delta \vbol \Vb}{\Rb} e^{\alpha } + \frac{\delta \vbol^2}{\Rb} e^{\alpha } - \delta P_{e-c} } \\[\verticaldistance]
        {\displaystyle \delta P_{e-c} - \delta P_{c-t} - \delta P_{c-s} }                                                                                       \\[\verticaldistance]
        {\displaystyle \delta P_{c-t} - \delta P_{t-s} }
    \end{bmatrix} + x.
\end{equation}
Specifically, $x = E_0 \delta(t)$ is an event with energy $E_0$, with
$\delta(t)$ being the Dirac delta. We use the SciPy \textsc{solve\_ivp}
\cite{2020SciPy-NMeth} numerical solver that takes in an initial condition such
that the crystal node starts with a change in temperature $t_c = E_0 /
    C_c(\overline{T}_c)$, with crystal heat capacity $C_c$ evaluated at equilibrium
temperature $\overline{T}_c$.

Since our heat capacities are temperature dependent, and they are sensitive to
the absolute temperature of the nodes, fitting these capacities is a method to
infer the absolute temperature of the nodes at equilibrium. Linearization of
the system around equilibrium can be simplified if we assume a constant
capacity matrix at equilibrium $C(T)$ (which easily holds in our system for the
equilibrium condition), and just taking the Jacobian of the right hand side
$f(\delta y)$, we get:
\begin{eqnarray}\label{eqn:lin_sys}
    C(T) \dot{\delta y} \ && \approx f(T) + \pdv{f_i}{(\delta y_j)}_{T}(\delta y)_j + x \nonumber \\
    && \equiv G(T) (\delta y) + x.
\end{eqnarray}
Note $f(T) = 0$ by the fact that at equilibrium the deviation
is 0. We will refer to this linear
system in the
noise analysis in \secref{sec:noise}.

\section{Equilibrium Model}\label{sec:eq_model}
In order to find an optimized operational resistance to achieve the best
signal-to-noise (SNR) ratio and stability at a given temperature, CUORE
incrementally tests the biasing voltage $V$ in \eqnref{eqn:electrical_1} and
records the NTD-Ge voltage and current.

For details of how the V-I curves are taken and how errors are propagated, we
refer to the work by Alfonso, et al \cite{alfonso2020}. The errors on the data
are limited by the 10\% accuracy of the bias resistor \cite{nutini2018}.

We use the equilibrium model to complement the dynamic model and to check the
consistency of our fitted parameters. We used the bottom diagram of
\figref{fig:thermal_circuit} as the thermal circuit, and solve the temperature
of each node in the thermal circuits \eqnsref{eqn:thermal_1} to
(\ref{eqn:thermal_3}) by setting the left hand side to 0 so that it is time
independent. The electron temperature is then used to calculate the NTD-Ge
resistance along with the NTD-Ge biasing voltage recorded in the V-I curves.
The calculated resistances are fitted to the V-I curve resistances by
minimizing the least squares between them.

During the CUORE optimization campaign, V-I curves at several temperatures
(from $\SI{12}{\milli \kelvin}$ to $\SI{27}{\milli \kelvin}$) were acquired
only for 78 of the 988 detectors, which were identified as representative
channels for the whole array. Our analyses in the rest of the paper apply to
these 78 channels. The NTD-Ge types in these towers cover all types used in
CUORE (tower 8 --- NTD 41C, tower 9 --- NTD 39C, tower 10 --- NTD 39D). As the
NTD-Ge thermistors in the same tower are diced from the same wafer, we make the
assumption that NTD-Ge parameters obtained from the same tower share common
parameter distributions, and consequently they can be used on a tower basis to
characterize the same type of NTD-Ge thermistors in other towers. We summarize
the fit results by NTD batches in \tabref{tab:lc_fit}. All NTD batches are
nominally irradiated to the same neutron fluence of about $\SI{3.4e18}{n \per
    \cm^2}$ at MIT Nuclear Research Laboratory, but the position with respect to
the neutron beam and the number of irradiation passes are different. The errors
in the table are standard deviations of the parameter spread in the same tower.

\begin{figure*}[!htb]
    \centering
    \subfloat[Channel 367 from tower 8.]{
    \includegraphics[width = 0.315\textwidth]{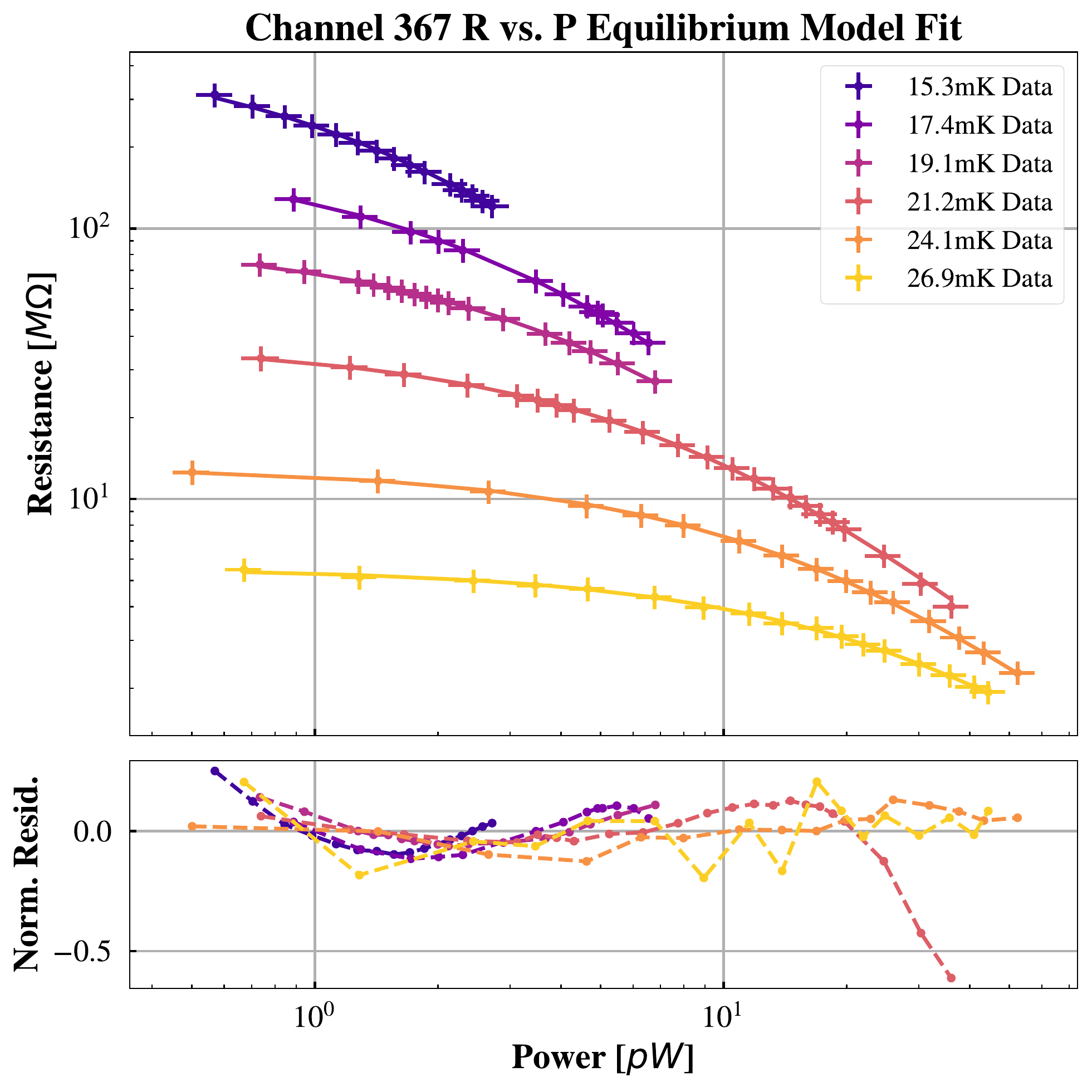}
    }
    \subfloat[Channel 419 from tower 9.]{
    \includegraphics[width = 0.315\textwidth]{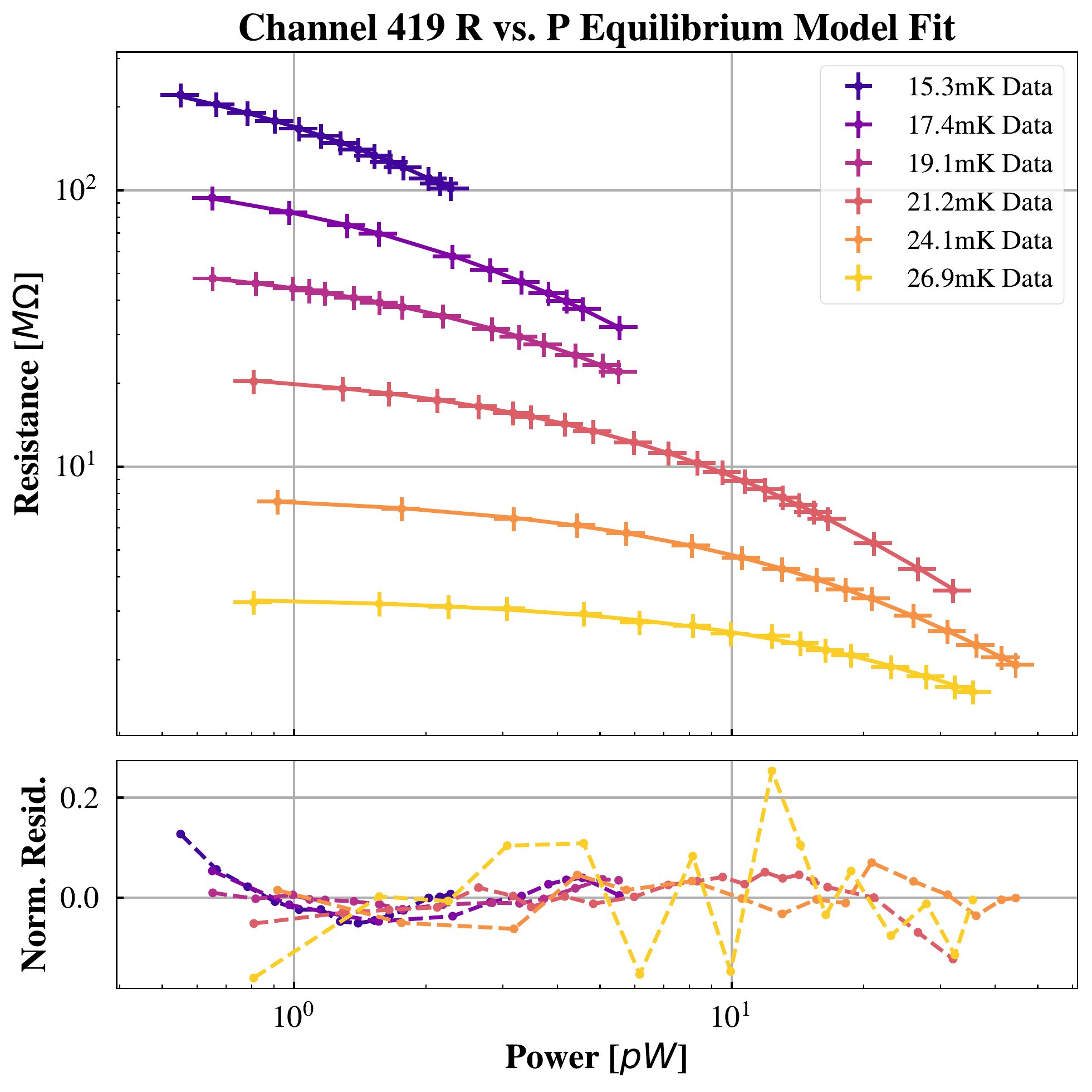}
    }
    \subfloat[Channel 487 from tower 10.]{
    \includegraphics[width = 0.315\textwidth]{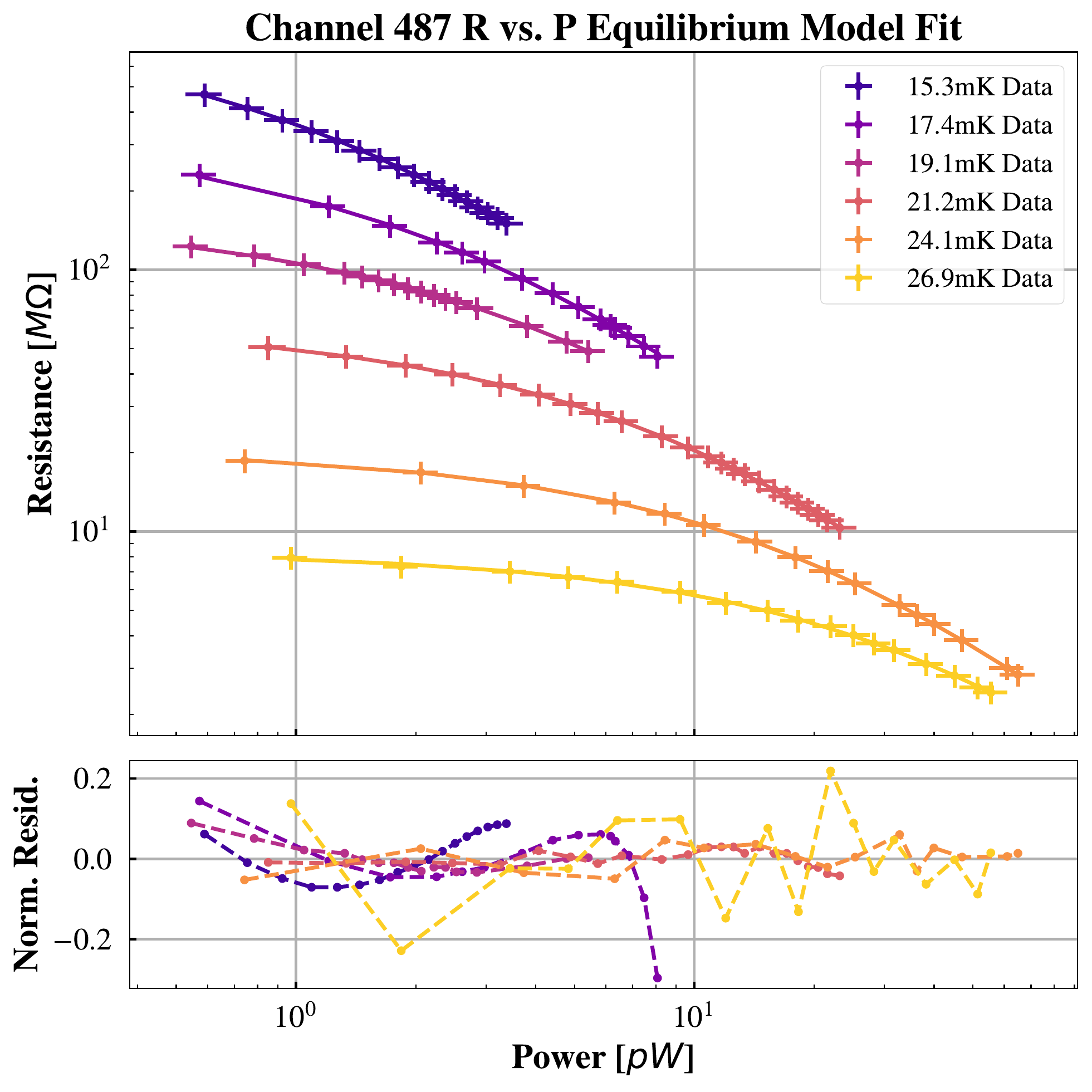}
    }
    \caption{V-I fits for three sample channels from tower 8, 9, and 10 of CUORE. Each tower represents a different batch of produced NTD-Ge. Each V-I curve was taken at different heat-sink temperatures. The electron-phonon
    coupling within the NTD-Ge thermistor is much weaker than other couplings between the calorimeter components.}
    \label{fig:lc_fit}
\end{figure*}
\begin{table*}[!htb]
    \centering
    \caption{Summary of NTD-Ge V-I curve fits by NTD-Ge batch.}
    \label{tab:lc_fit}
    \adjustbox{width=\textwidth, center=\textwidth}{
    \begin{tabular}{CCCCCCc}
        \noalign{\vskip 1mm}
        \hline
        \hline
        \noalign{\vskip 1mm}
        P_{\textrm{os}} \ [\si{\watt}]             & R_0  \ [\si{\ohm}] & T_0 \ [\si{\kelvin}] & a_{e-c}       & g_{e-c} \ [\si{\watt / \kelvin^{a_{e-c}}}] & \lambda_0 \ [\si{\nano \meter \cdot \kelvin^{0.5}}] & NTD batch \\[0.5ex]
        \hline
        \noalign{\vskip 1mm}
        \left(7.70 \pm 0.62\right) \times 10^{-13} & 0.474 \pm 0.035    & 7.04 \pm 0.19        & 5.36 \pm 0.13 & 0.0154 \pm 0.0082                          & 13.8 \pm 2.6                                        & 39C       \\
        \left(7.8 \pm 1.5\right) \times 10^{-13}   & 0.58 \pm 0.15      & 7.58 \pm 0.15        & 5.81 \pm 0.42 & 0.11 \pm 0.15                              & 7.9 \pm 7.6                                         & 39D       \\
        \left(6.90 \pm 0.80\right) \times 10^{-13} & 0.622 \pm 0.053    & 6.78 \pm 0.16        & 5.70 \pm 0.18 & 0.042 \pm 0.020                            & 3.8 \pm 4.3                                         & 41C       \\
        \hline
        \hline
    \end{tabular}
    }
\end{table*}
We have observed that for the electron-phonon
coupling
within the NTD-Ge thermistor, its thermal conductance G is much smaller
than the thermal conductances for coupling between the
$\text{TeO}_{2}$ crystal
and the heat-sink, and general coupling between the PTFE support and the
heat-sink.
This observation implies that the V-I data alone could not determine the
power law of thermal
coupling beyond that of the electron-phonon coupling. The fitting is
not sensitive to changes in initial values for coupling between the
$\text{TeO}_{2}$ crystal and the heat-sink
or the coupling between the PTFE support and the heat-sink.
We plot the fits on sample channels in \figref{fig:lc_fit}.

We also note that the $T_0$ values, as appearing in \eqnref{eqn:ohmic_vrh}, are
about one to two Kelvins higher than previously reported measurements
\cite{Rusconi:2011xud}, while the $R_0$ values are lower than previous
characterizations. The $R_0$ parameter is highly anti-correlated with $T_0$ by
the nature of the \eqnref{eqn:ohmic_vrh}, which should explain the fits'
underestimation. The electron-phonon coupling power law coefficient $a_{e-c}$,
and $\lambda_{0}$ are close to expectation: $a_{e-c}$ values are between 5 and
6, while $\lambda_0$ parameters \cite{piat2001} are on the order of
$\SI{10}{\nano \meter \cdot \kelvin^{0.5}}$.

\section{Dynamic Model}\label{sec:dyn_model}
The dynamic model aims to explain the energy dependence of calorimeter signal
pulses. A linear system has a constant conductivity matrix $G(T)$ and a
constant capacity matrix $C(T)$ in \eqnref{eqn:lin_sys}, and thus it does not
predict any change in pulse shape regardless of the input energy because
eigenvalues of the these two matrices are fixed. This is not true for CUORE
pulses, and we seek to model the detector response using \eqnref{eqn:ode_sys}
for pulses up to the alpha region ($< 6$ MeV) in this section.

We use the same model as we have verified using the V-I curves in
\secref{sec:eq_model}. For a given base temperature and NTD working point, we
group a set of 5 pulses with different energies together and perform a
simultaneous fit on each group of 5 pulses to increase our model sensitivity
towards energy dependence of the pulse shape. Each group includes energies
ranging from $\SI{511}{\kilo \electronvolt}$ to $\SI{5407}{\kilo
    \electronvolt}$, which are chosen from the energy spectrum peaks of the
dataset. We input these energies as the initial condition for the simulations,
and fit by minimizing the squared difference between the raw pulses and
simulations. We are also fitting the heat-sink temperature, assuming that the
heat-sink temperatures for all of the pulses are close enough so that the five
pulses can be described by the same heat-sink temperature. The model is tested
on the same 78 channels as introduced in \secref{sec:eq_model}. Each channel
has 40 "groups" of simultaneous fits. To correct for signal baseline drifts, we
have added two corrections for each pulse: a linear correction parametrized by
$V_{\rm{lin}} = m (t - b)$, where $m$ is the slope and $b$ is the intercept;
and an exponential tail correction parametrized by $V_{\rm{exp}} = A e^{-t /
        \tau}$, where $A$ is the initial amplitude at time 0 and $\tau$ is the
characteristic decay time. The linear correction is often caused by thermal
drifts of the detector or changes of the cryostat's noise environment,
especially vibration noises at low frequencies, whereas the exponential
correction usually results from the decay of a previous event sufficiently
close to the examined one. The fit results for key model parameters are listed
in \tabref{tab:fit_pars}.
\begin{table*}[!htb]
    \centering
    \caption{Table of key parameter initial values used in the fit. If the parameter has an error estimation
    in the table, then a parameter is given a prior. The prior is
    assumed to be gaussian and given a boosted weighting of 100, as otherwise
    we find it hard for the priors to be effective in constraining the fit
    parameters. The parameter expectation values are confined by previous
    measurements \cite{pedretti2004, vignati2009, alessandrello1993, carrettoni2011}.}
    \label{tab:fit_pars}
    \adjustbox{width=\textwidth, center=\textwidth}{
    \begin{tabular}{lccc}
        \noalign{\vskip 1mm}
        \hline
        \hline
        \noalign{\vskip 1mm}
        Parameter Name                                         & Symbol            & Unit                                    & Value                                    \\[0.5ex]
        \hline
        \noalign{\vskip 1mm}
        Pulse Time                                             & $ t_{\textrm{p}}$ & $\si{\second}$                          & $ 3$                                     \\
        Parasitic Capacitance                                  & $ c_{\textrm{p}}$ & $\si{\pico \farad}$                     & $ \num{500\pm100} $                      \\
        Electron Heat Capacity \cite{nutini2018, Alduino_2016} & $ c_{e}$          & $\si{\joule / \kelvin}$                 & $ \num{8.6e-11} $                        \\
        $\text{TeO}_{2}$ Crystal Heat Capacity                 & $ c_{c}$          & $\si{\joule / \kelvin^4}$               & $ \num{2.2\pm0.2 e-3} $                  \\
        PTFE Support Heat Capacity                             & $ c_{t}$          & $\si{\joule / \kelvin^2}$               & $ \num{2.1\pm0.2 e-6} $                  \\
        Electron-Phonon Thermal Conductivity                   & $ g_{e-c}$        & $\si{\watt / \kelvin^{a_{e-c}}}$        & From \tabref{tab:lc_fit} (NTD-dependent) \\
        Crystal-PTFE Thermal Conductivity                      & $ g_{c-t}$        & $\si{\watt / \kelvin^{a_{c-tef}}}$      & $ \num{1e-8}$                            \\
        PTFE-heat-sink Thermal Conductivity                    & $ g_{t-s}$        & $\si{\watt / \kelvin^{a_{tef-s}}}$      & $ \num{1e-8}$                            \\
        Electron-Phonon Conductivity Power Exponent            & $ a_{e-c}$        & N/A                                     & From \tabref{tab:lc_fit} (NTD-dependent) \\
        Absorber-PTFE Conductivity Power Exponent              & $ a_{c-t}$        & N/A                                     & $1$ (Fixed)\footnotemark[1]              \\
        PTFE-hea-sink Conductivity Power Exponent              & $ a_{t-s}$        & N/A                                     & $1$ (Fixed)\footnotemark[1]              \\
        NTD-Ge Characteristic Temperature                      & $ T_{0}$          & $\si{\kelvin}$                          & From \tabref{tab:lc_fit} (NTD-dependent) \\
        NTD-Ge Characteristic Hopping Length                   & $ \lambda_0$      & $\si{\nano \meter \cdot \kelvin^{0.5}}$ & From \tabref{tab:lc_fit} (NTD-dependent) \\
        Heat-sink Temperature                                  & $ T_{\rm{base}}$  & $\si{\kelvin}$                          & $ 0.010 \sim 0.015$                      \\
        \hline
        \hline
    \end{tabular}
    \footnotetext[1]{We have tested floating the power exponent around 3 in the
    fit: it does not impact the least square loss.
    Thus we have fixed the exponential to 1 to demonstrate that the pulse shape
    is not sensitive to these thermal couplings.}
    }
\end{table*}

\begin{figure}[!htb]
    \centering
    \includegraphics[width = 0.45\textwidth]{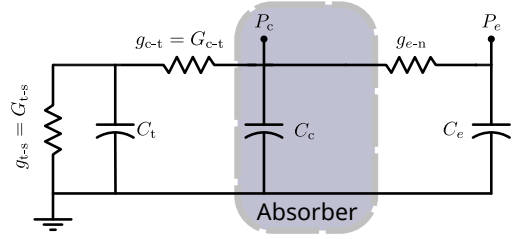}
    \caption{The thermal circuit used in the dynamic model. The direct coupling between the crystal and heat-sink has been removed because the fit is not sensitive to this parameter.}
    \label{fig:dyn_thermal_circuit}
\end{figure}

During the fitting process, we find that despite the use of simultaneous fit,
the least square loss converges slowly if too many input parameters are
provided. Therefore, we set the thermal coupling between the $\text{TeO}_{2}$
crystal and the heat-sink to zero, and have also fixed both power exponents of
the coupling between the crystal and its PTFE support and the coupling between
the PTFE support and the heat-sink to 1 so that the heat conductance becomes
independent of temperature. Both changes have no noticeable effects on the fit
results. The equivalent thermal circuit diagram is shown in figure
\figref{fig:dyn_thermal_circuit}.

\begin{figure}[!htb]
    \centering
    \includegraphics[width = 0.45\textwidth]{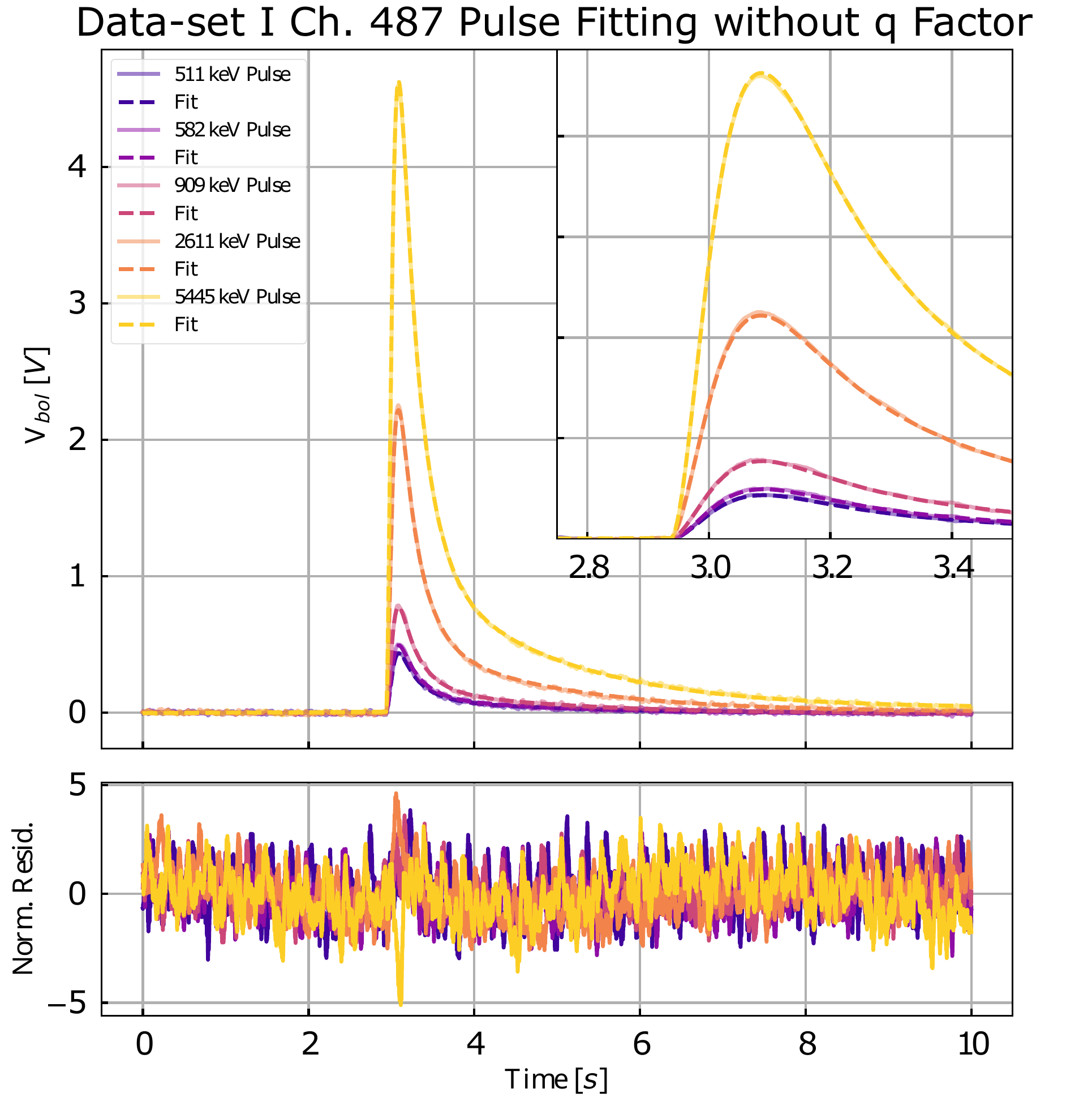}
    \caption{Grouped pulse fit for sample channel 487 for data-set I ($\SI{15.3}{\milli \kelvin}$). The residual normalized with the standard deviation of the first 2 seconds of the waveform is plotted at the bottom. The inset shows a zoomed portion of the peak. }
    \label{fig:pulse_fit_base_3522}
\end{figure}

\begin{figure}[!htb]
    \centering
    \includegraphics[width = 0.45\textwidth]{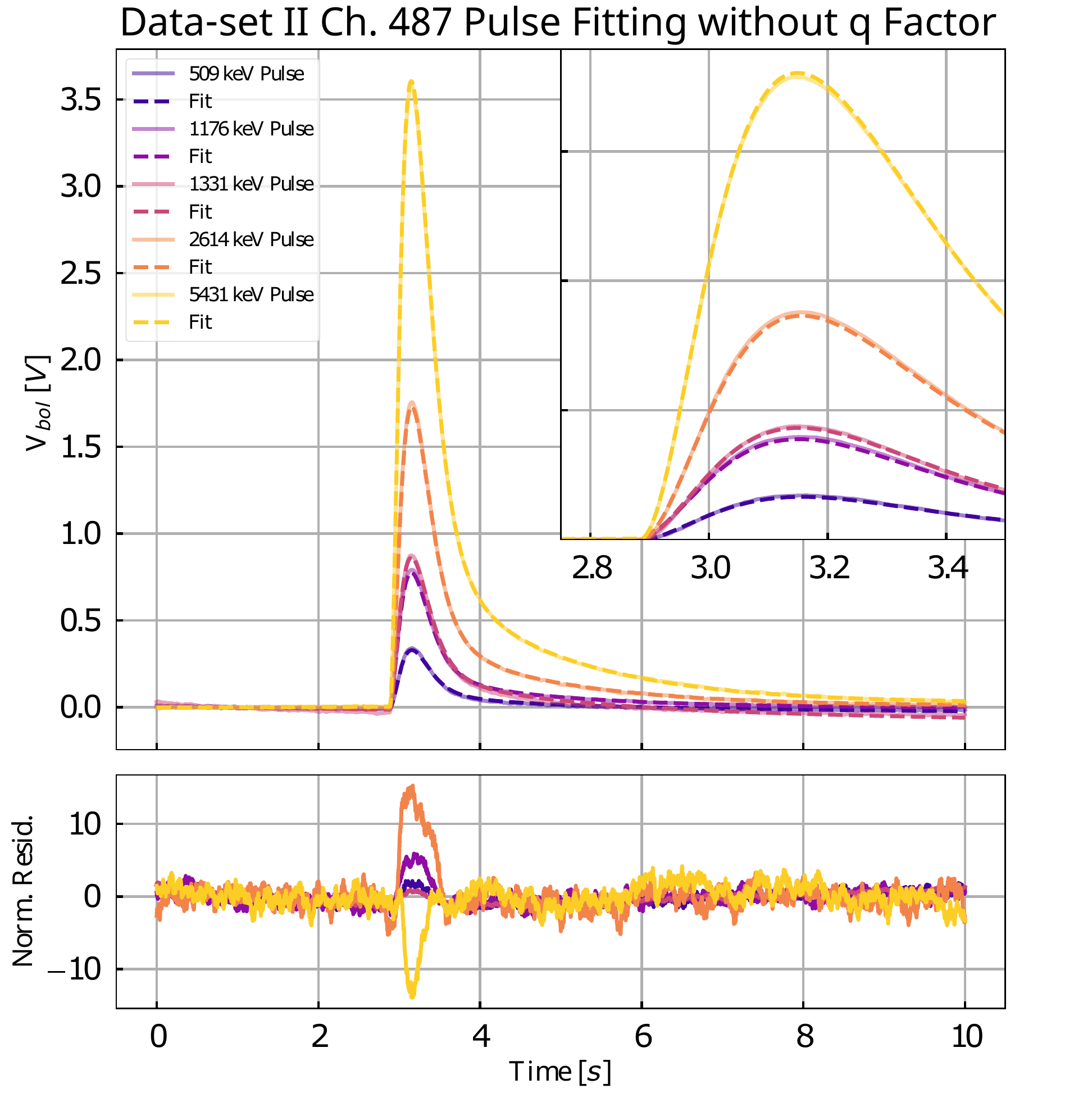}
    \caption{Grouped pulse fit for sample channel 487 for data-set II ($\SI{11.9}{\milli \kelvin}$). The normalized residuals are large near the peak of the signal pulses. This means the model pulse is over-shooting the peak in the highest energy, while the other pulses are underestimating the peak. The model without the $q$ factor cannot address this deviation. }
    \label{fig:pulse_fit_base_3564}
\end{figure}
\figref{fig:pulse_fit_base_3522} (for data-set I, $\SI{15.3}{\milli \kelvin}$) and \figref{fig:pulse_fit_base_3564}
(for data-set II, $\SI{11.9}{\milli \kelvin}$) both show one of the sample
group
fits. To compare with the equilibrium model, we tabulate
the mean and the standard deviation of
key fit parameters in \tabref{tab:fit_pars_base_result}.
We note that almost all key parameters from the V-I curve
model are close to the fitting results from the dynamic model,
suggesting that the dynamic model is consistent with the earlier
equilibrium model.
\begin{table*}[!htb]
    \centering
    \caption{
    Principal fit parameters of the dynamic model, grouped by NTD-Ge batch. The Roman numeral subscript after $T_s$ denotes the data-set number. The parameter fit values for both data-sets are compatible with each other and thus combined.}
    \label{tab:fit_pars_base_result}
    \adjustbox{width=\textwidth, center=\textwidth}{
    \begin{tabular}{CCCCCCCc}
        \noalign{\vskip 1mm}
        \hline
        \hline
        \noalign{\vskip 1mm}
        c_\textrm{p} \ [\si{\pico \farad}] & g_{e-c} \ [\si{\watt / \kelvin^{a_{e-c}}}] & a_{e-c}       & T_0 \ [\si{\kelvin}] & \lambda_0 \ [\si{\nano \meter \cdot \kelvin^{0.5}}] & T_{s, I} \ [\si{\milli \kelvin}] & T_{s, II} \ [\si{\milli \kelvin}] & NTD batch \\[0.5ex]
        \hline
        \noalign{\vskip 1mm}
        511 \pm 17                         & 0.026 \pm 0.002                            & 5.07 \pm 0.07 & 7.039 \pm 0.009      & 12.3 \pm 0.9                                        & 13.1 \pm 0.8                     & 12.2 \pm 1.4                      & 39C       \\
        501 \pm 9                          & 0.17 \pm 0.02                              & 5.52 \pm 0.10 & 7.576 \pm 0.004      & 8.6 \pm 4.2                                         & 13.0 \pm 0.9                     & 11.3 \pm 0.6                      & 39D       \\
        514 \pm 14                         & 0.058 \pm 0.002                            & 5.34 \pm 0.06 & 6.771 \pm 0.007      & 2.0 \pm 1.3                                         & 13.3 \pm 0.4                     & 11.3 \pm 0.6                      & 41C       \\
        \hline
        \hline
    \end{tabular}
    }
\end{table*}

\begin{figure}[!htb]
    \centering
    \includegraphics[width = 0.45\textwidth]{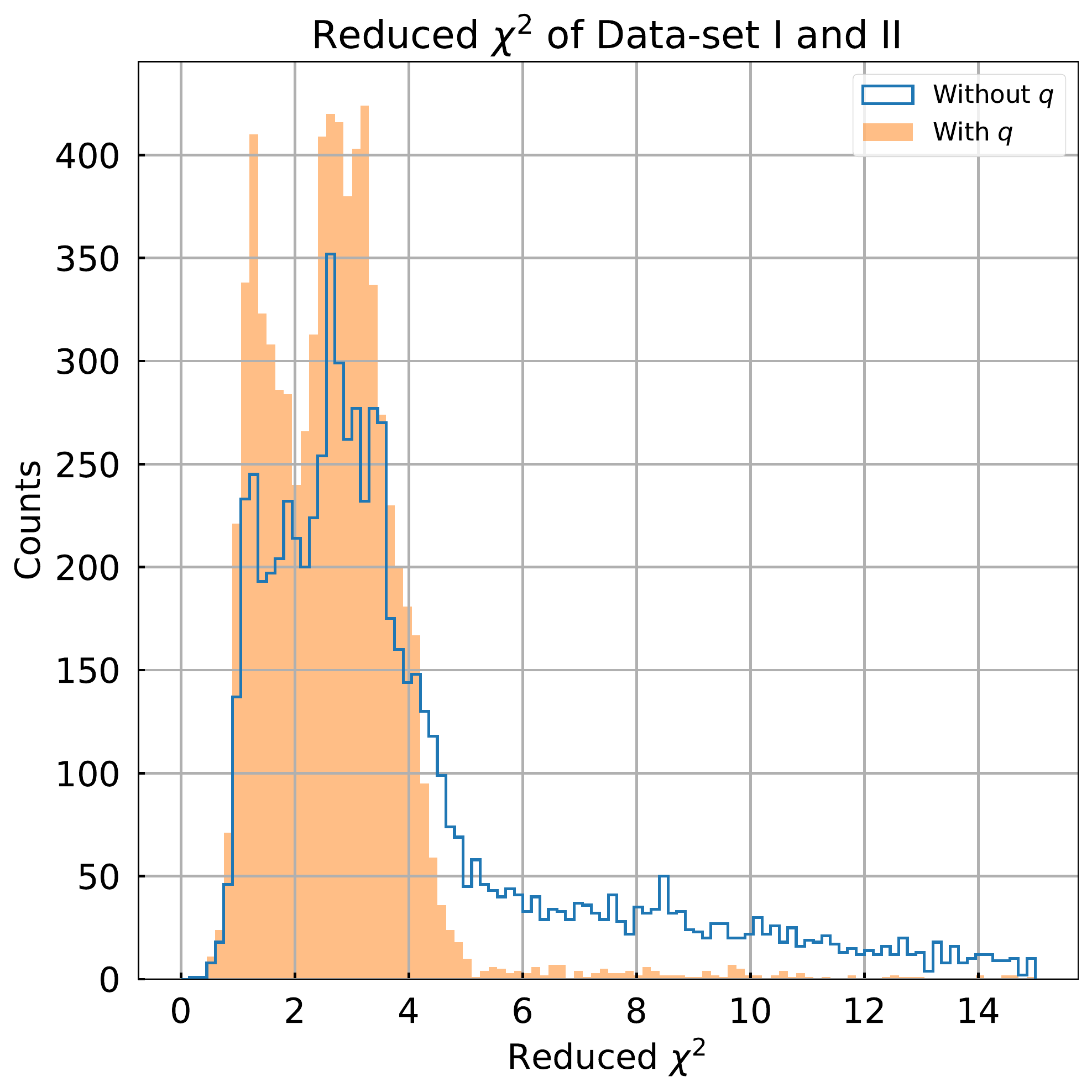}
    \caption{Comparison between the histograms of the reduced $\chi^2$ values for the simultaneous fits performed on data-sets I and II. The higher reduced $\chi^2$ peak in the "double-bump" feature corresponds to channels in the NTD batches 41C and 39C, which in general show a poorer quality of fit even without the $q$ factor.}
    \label{fig:red_chi2_comp}
\end{figure}

\begin{figure}[!htb]
    \centering
    \includegraphics[width = 0.45\textwidth]{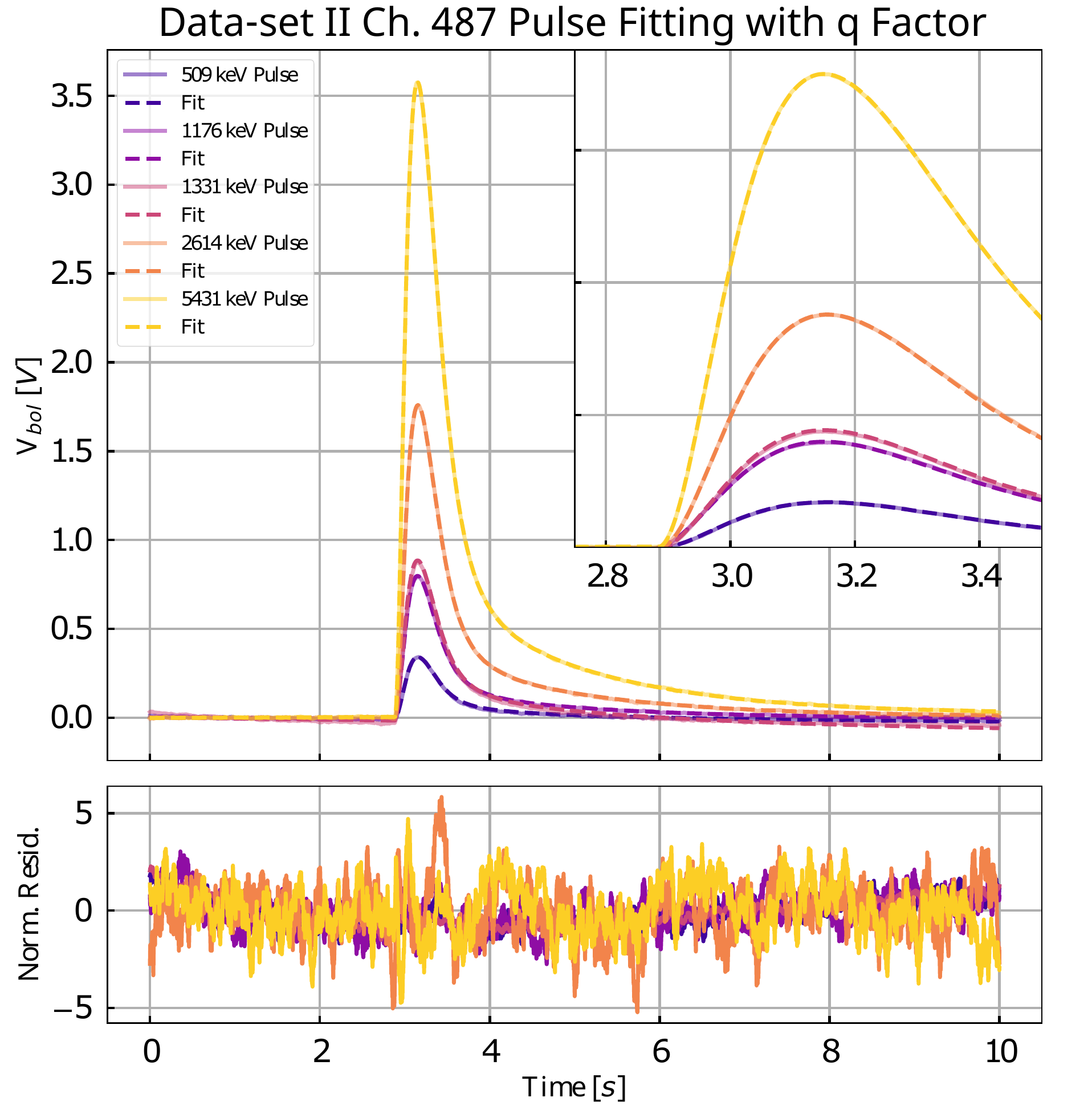}
    \caption{Grouped pulse fit for sample channel 487 for data-set II with $q$ factor. Cf. \figref{fig:pulse_fit_base_3564}. The normalized residuals for all the pulses are closer to being evenly distributed around 0. }
    \label{fig:pulse_fit_base_q_3564}
\end{figure}

\begin{table*}[!htb]
    \centering
    \caption{Principal fit parameters of the dynamic model with $q$ factor, grouped by NTD batch. }
    \label{tab:fit_pars_base_q_result}
    \adjustbox{width=\textwidth, center=\textwidth}{
    \begin{tabular}{CCCCCCCCc}
        \noalign{\vskip 1mm}
        \hline
        \hline
        \noalign{\vskip 1mm}
        c_{\textrm{p}} \ [\si{\pico \farad}] & g_{e-c} \ [\si{\watt / \kelvin^{a_{e-c}}}] & a_{e-c}       & T_0 \ [\si{\kelvin}] & \lambda_0 \ [\si{\nano \meter \cdot \kelvin^{0.5}}] & q          & T_{s, I} \ [\si{\milli \kelvin}] & T_{s, II} \ [\si{\milli \kelvin}] & NTD batch \\[0.5ex]
        \hline
        \noalign{\vskip 1mm}
        513 \pm 18                           & 0.026 \pm 0.002                            & 5.05 \pm 0.07 & 7.039 \pm 0.008      & 12.4 \pm 0.9                                        & -34 \pm 15 & 12.8 \pm 0.9                     & 10.7 \pm 1.2                      & 39C       \\
        507 \pm 10                           & 0.17 \pm 0.02                              & 5.48 \pm 0.10 & 7.577 \pm 0.004      & 10.5 \pm 3.8                                        & -32 \pm 11 & 12.7 \pm 1.0                     & 10.6 \pm 1.2                      & 39D       \\
        523 \pm 15                           & 0.059 \pm 0.002                            & 5.34 \pm 0.06 & 6.770 \pm 0.007      & 2.7 \pm 1.4                                         & -32 \pm 10 & 13.2 \pm 0.4                     & 11.1 \pm 0.7                      & 41C       \\
        \hline
        \hline
    \end{tabular}
    }
\end{table*}
However, the residuals stray away from zero near the peak of
the pulses. We find that introducing an empirical second-order correction term
to the NTD-Ge resistivity expansion, defined in \eqnref{eqn:resistance_expansion},
produces better fitting results. To test if the additional second-order
correction can be explained by the
standard law, i.e. \eqnref{eqn:electrical_vrh}, we separate its expansion
to the second order so that:
\begin{eqnarray}
    \alpha \to \alpha && - \frac{3}{4} \eta \left(\frac{\delta t_{e}}{T_{e}} \right)^2 + \frac{15}{8} C_{\lambda} \frac{\Vb}{T_{e}^{1.5}} \left(\frac{\delta t_{e}}{T_{e}} \right)^2 \nonumber \\
    && - \frac{3}{2} C_{\lambda} \frac{\Vb}{T_{e}^{1.5}} \left(\frac{\delta t_{e}}{T_{e}} \right) \left(\frac{\delta \vbol}{\Vb} \right) + q \left(\frac{\delta t_{e}}{T_{e}} \right)^2.
\end{eqnarray}
Our null hypothesis is that the additional $q$ factor is 0.
Fit results on data-sets I and II with the inclusion of
the $q$ factor have notably improved, as can be
inferred from the comparison of the histograms of
reduced $\chi^2$ figure-of-merit in \figref{fig:red_chi2_comp}.
\figref{fig:pulse_fit_base_q_3564} showing the fitting of the same group of pulses
as those in \figref{fig:pulse_fit_base_3564} confirms this observation.

The parameters in the modified dynamic model
are summarized in \tabref{tab:fit_pars_base_q_result},
where we have applied a reduced $\chi^2$ cut of 5 in the fitting
results for
filtering out spurious pulses when calculating the statistics. Pulses with high
reduced $\chi^2$ could be caused by unstable baseline or
pileups. The cutoff is supposed to exclude these not well-fitted pulses
so that the fit parameters are more credible.

As listed in the table, the $q$ factors have distributions not compatible with
the null hypothesis and are negative. This result suggests possible unaccounted
correction terms in the NTD-Ge characteristic function. It is not likely that
such factor arises from subtleties in the thermal couplings between the crystal
and the PTFE support or the heat-sink, as they affect the long pulse decay time
constant. Since a negative $q$ factor means that the expansion $\alpha$ is more
concave down, it could implie a larger $\gamma$ value in
\eqnref{eqn:ohmic_vrh}. More experimental efforts are necessary in the future
to determine the origin of this correction.

At the end of this section, by comparing to the equilibrium model results in
\tabref{tab:lc_fit}, we note that most values we obtain in
\tabref{tab:fit_pars_base_q_result} are consistent. The heat-sink temperature
fitting results have systematic deviations of around $\si{-1}$ to
$\SI{-2.5}{\milli \kelvin}$. Another difference is that the power law exponents
$a_{e-c}$ are universally lower in the dynamic model than in the equilibrium
model. This implies that other thermal couplings with smaller power exponents
may be present in the system.

\section{Noise}\label{sec:noise}

Obtaining the physical parameters allows us to analyze major noise
contributions from both thermal and electrical circuits. The numerical
simulation can be carried out in the frequency domain. We begin by performing a
Fourier transformation on the linear system \eqnref{eqn:lin_sys}:
\begin{equation}\label{eqn:lin_sys_freq}
    ( i\omega C-G) Y( \omega ) =X( \omega ).
\end{equation}
The right hand side of the equation represents the noise input, and with
appropriate input vector the output vector $Y$ simulates
the
detector response.

There are two major kinds of noises: Thermal Fluctuation Noise (TFN) and
Johnson noise from the readout circuit. Both have frequency-independent power
spectra, but TFN has a stronger temperature dependency. We can use the TFN
between charge carriers and NTD-Ge lattice as an example of incorporating TFN.
The noise source features the following spectrum \cite{mccammon2005}:
\begin{equation}
    p^{2}_{T-e-n} =4kT^{2} GF_{\textrm{link}},
\end{equation}
where
\begin{equation}
    F_{\textrm{link}} =\frac{( T_{e} /T_{n})^{\beta +2} +1}{2},
\end{equation}
and where $\beta$ is the temperature dependence of
$G$ and we note here that
$P\sim gT^{\beta+1}$.
For TFN between two nodes at equilibrium with each other, the link term
$F_{\textrm{link}} = 1$.
In our system (radiative limit), the link term accounts for
the temperature gradient at the two ends.

Then our input vector is:
\begin{equation}
    X( \omega ) =\begin{bmatrix}
        0          \\
        p_{T-e-c}  \\
        -p_{T-e-c} \\
        0
    \end{bmatrix}.
\end{equation}
We note that an
increase of power for the charge carriers in the context of
noise indicates a decrease of power on the opposite side, and hence the minus
sign before $p_{T-e-c}$. The absolute value
of the output vector is the average noise power due to electron-phonon TFN.

\begin{figure}[!htb]
    \centering
    \includegraphics[width = 0.45\textwidth]{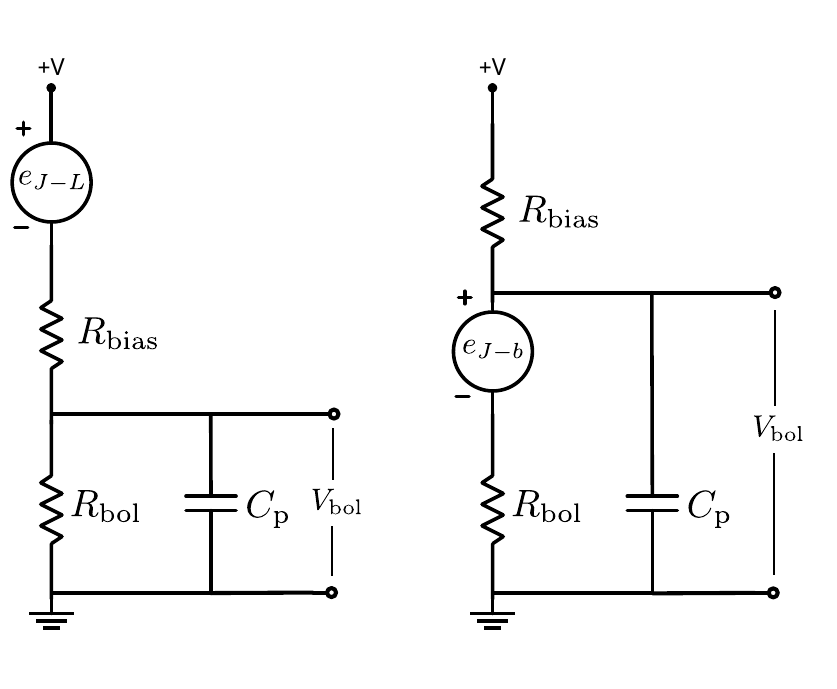}
    \caption{Equivalent Johnson noise generation circuits for the bias resistor $\RL$ and NTD-Ge thermistor $\Rb$.}
    \label{fig:noise_circuits}
\end{figure}

To evaluate the contribution of the Johnson noise, there are two parts: the
bias resistor and the thermistor. We refer to a similar study on
transition-edge sensors (TES) for the derivation \cite{figueroa-feliciano2006}.
Noise from the bias resistor can be expressed as an external voltage source, as
shown in \figref{fig:noise_circuits}.

We emphasize that the physical dimension of the input $X$ is equivalent to $GY$
and is in $\si{\ampere / \sqrt{\hertz}}$ for the electrical node, so we need to
use the noise current (the voltage divided by resistance) instead of voltage:
\begin{equation}
    i_{J-L} = \frac{e_{J-L}}{\RL}.
\end{equation}
In general the linearized system takes care of the electro-thermal
feedback through the matrix $G$, and thus the input is:
\begin{equation}
    X( \omega ) =\begin{bmatrix}
        i_{J-L} \\
        0       \\
        0       \\
        0
    \end{bmatrix}.
\end{equation}

Regarding the Johnson noise for NTD-Ge, we refer to the right plot of
\figref{fig:noise_circuits} for the noise circuit. The effective input on the
electrical node is still current input, but this time there is an additional
power correction $-\overline{I}_{b} v_{J-b}$ as the measured voltage includes
the voltage across the thermistor (responsible for electro-thermal feedback)
plus the voltage noise. The input vector is then:
\begin{equation}
    X( \omega ) =\begin{bmatrix}
        i_{J-b}                   \\
        -\overline{I}_{b} v_{J-b} \\
        0                         \\
        0
    \end{bmatrix}.
\end{equation}

Incorporating other thermal fluctuation noises, we find the noise input vector
as the following:
\begin{equation}
    X( \omega ) =\begin{bmatrix}
        i_{J-L} +i_{J-b}                     \\
        -\overline{I}_{b} v_{J-b} +p_{T-e-c} \\
        -p_{T-e-c} + -p_{T-c-tef}            \\
        p_{T-c-tef} + p_{T-tef-s}
    \end{bmatrix}.
\end{equation}

\begin{figure*}[!htb]
    \centering
    \subfloat[Channel 487, data-set I.]{
    \label{fig:base_nps_DS3522_487}
    \includegraphics[width = 0.31\textwidth]{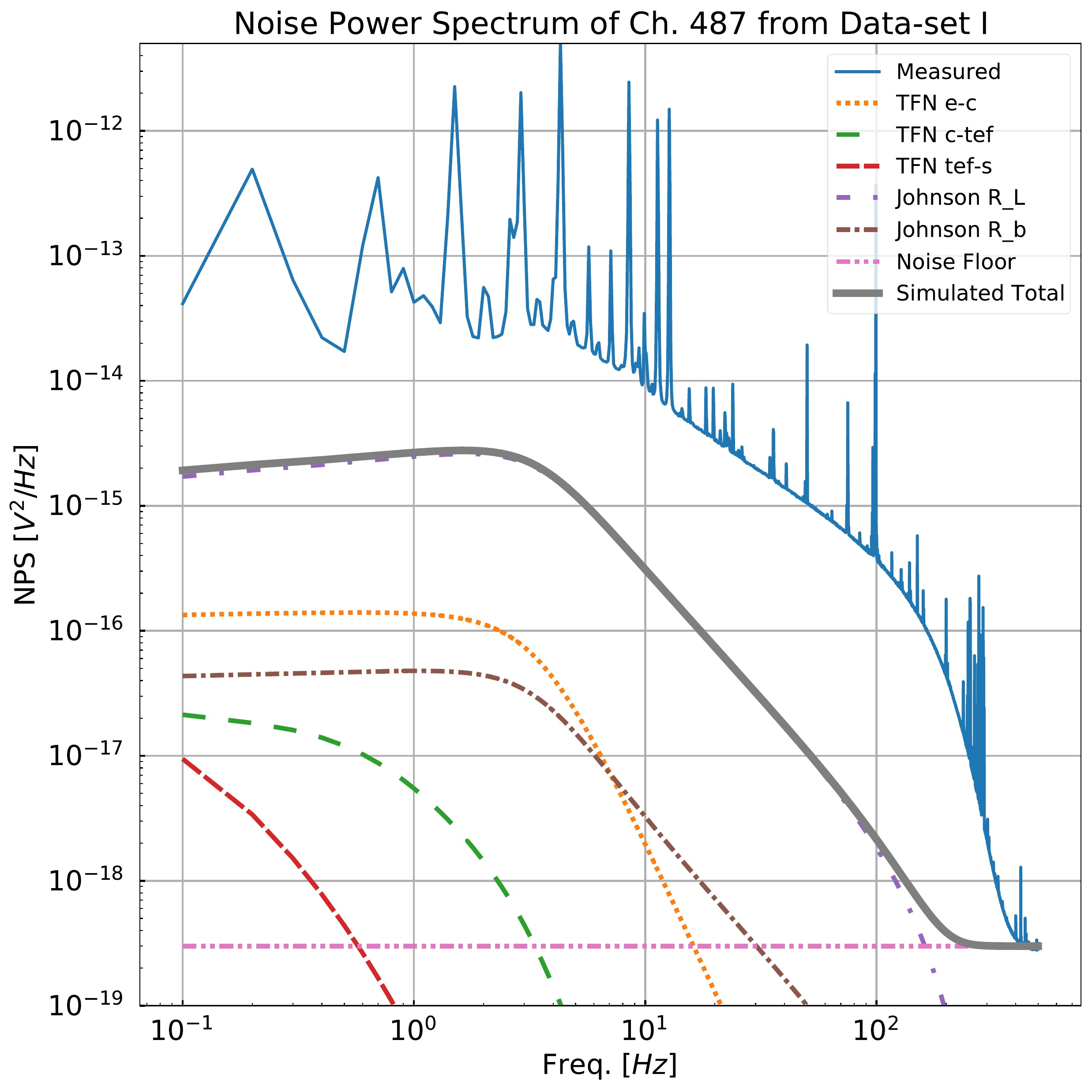}
    }
    \subfloat[Channel 487, data-set II.]{
    \label{fig:base_nps_DS3564_487}
    \includegraphics[width = 0.31\textwidth]{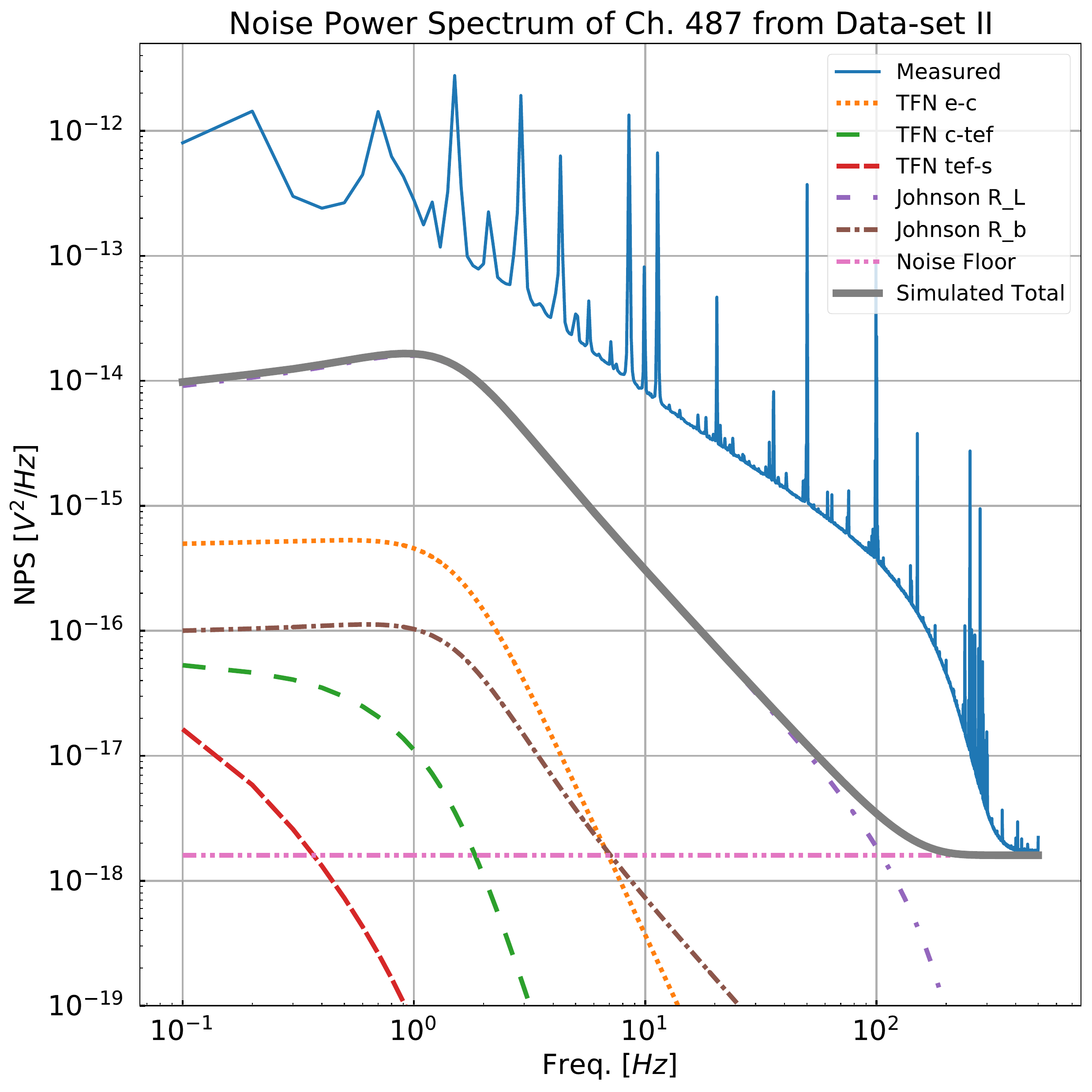}
    }
    \subfloat[Channel 474, data-set II.]{
    \label{fig:base_nps_DS3564_474}
    \includegraphics[width = 0.31\textwidth]{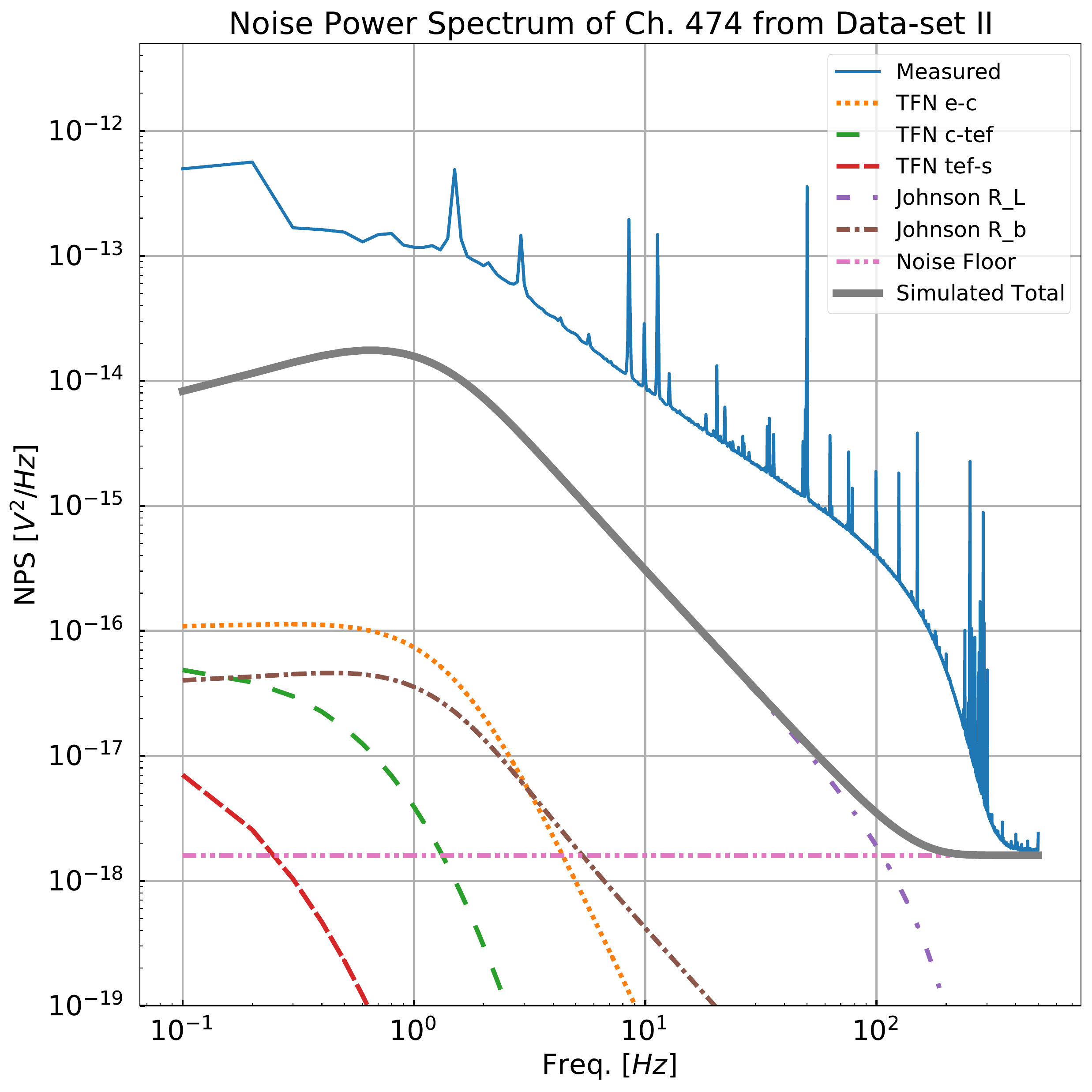}
    }
    \caption{Left and Middle: Simulated NPS for channel 487 overlaid with measured NPS from data-set I and II. Right: NPS for channel 474 from data-set II. We include this channel because it is more representative of an average CUORE channel. Spectra are divided by the gain of the amplification chain and multiplied with the Bessel filter transfer function. }
\end{figure*}

\begin{figure*}[!htb]
    \centering
    \subfloat[Channel 487, data-set I.]{
    \label{fig:bad_nps_DS3522}
    \includegraphics[width = 0.45\textwidth]{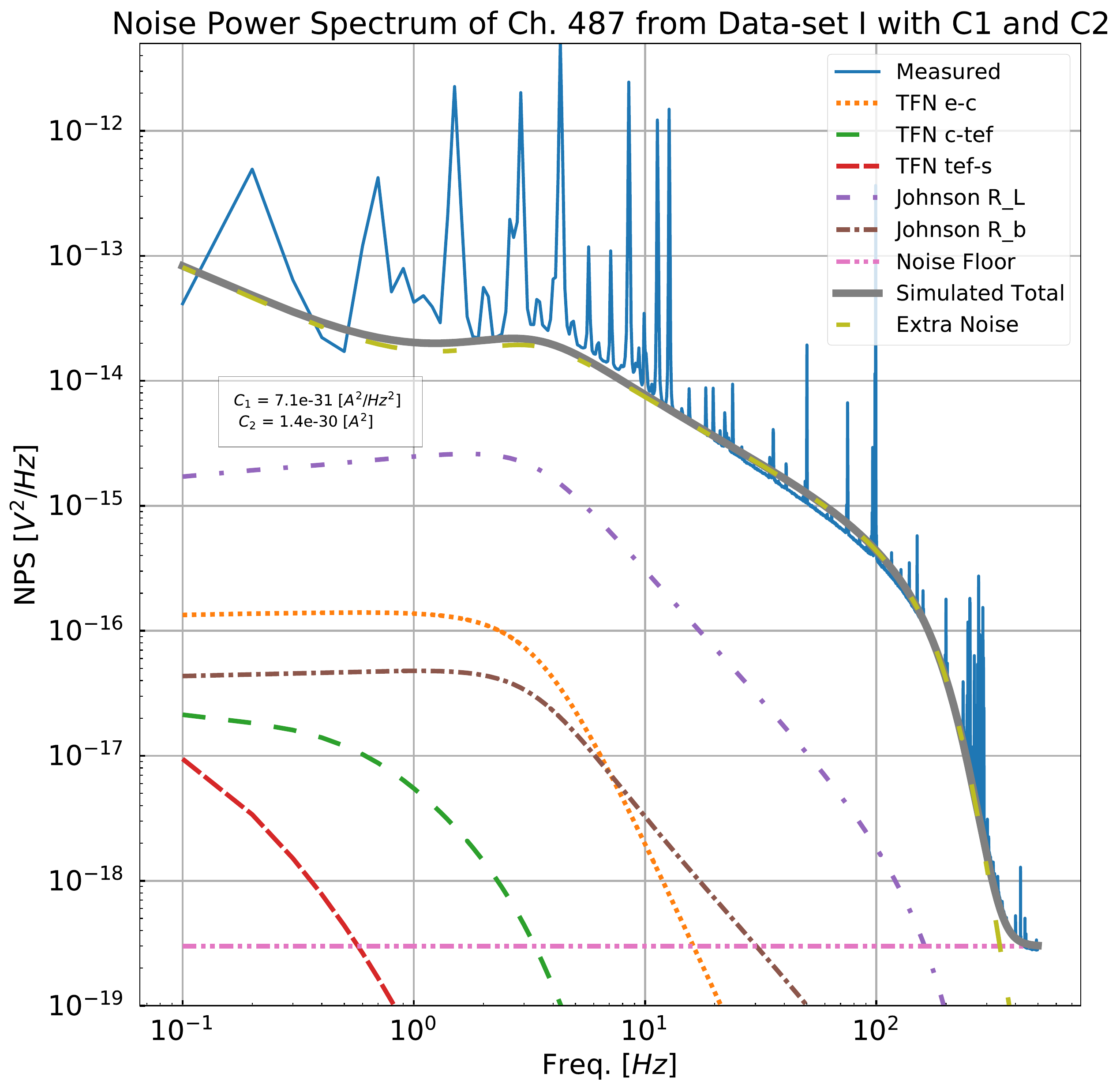}
    }
    \subfloat[Channel 487, data-set II.]{
    \label{fig:bad_nps_DS3564}
    \includegraphics[width = 0.45\textwidth]{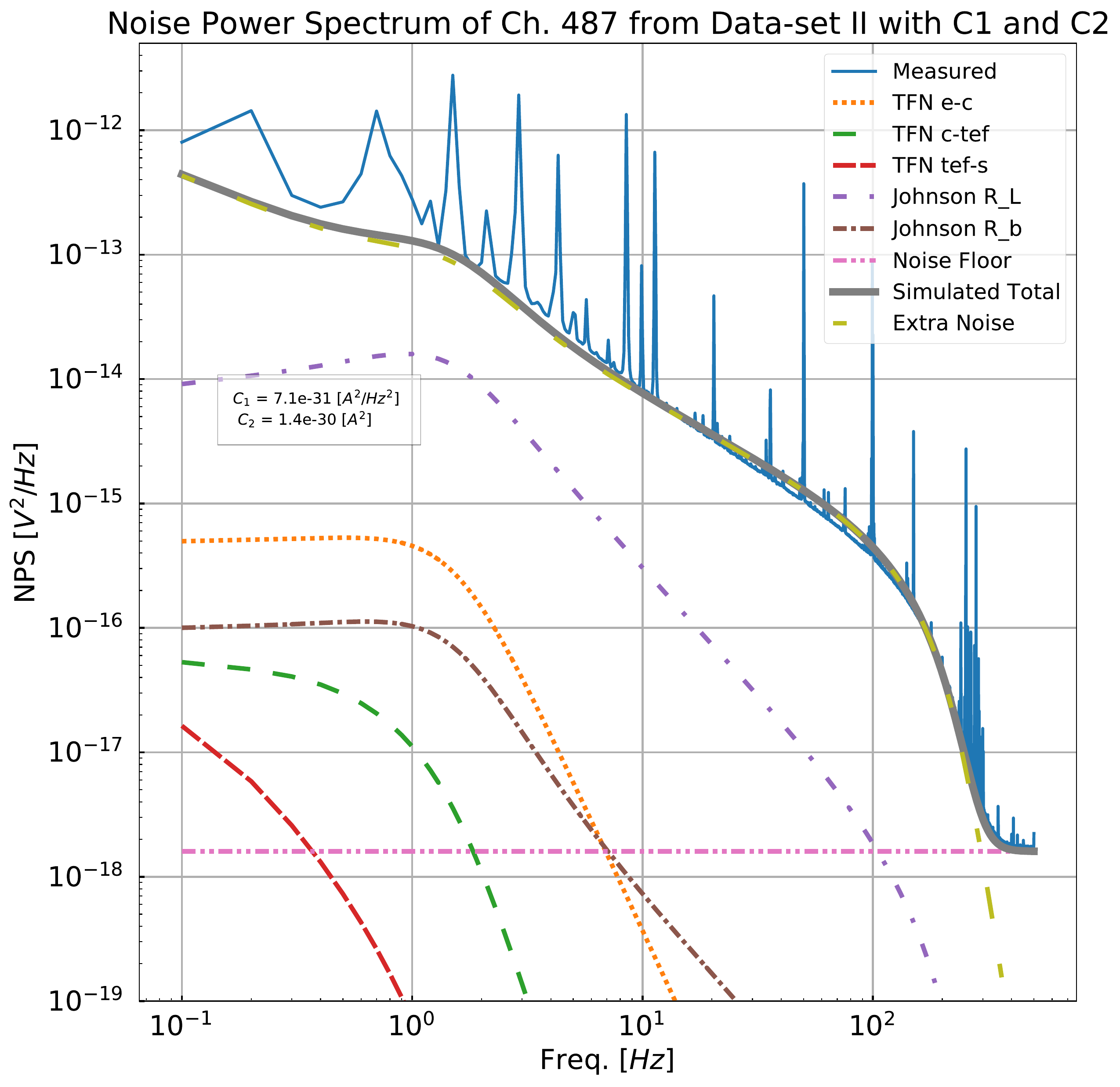}
    }
    \caption{Left: Simulated NPS with estimated $C_1$ and $C_2$ coefficients for the sample channel 487, compared to measured NPS from data-set I. Right: Simulated NPS with the same estimated $C_1$ and $C_2$ coefficients, compared to measured NPS from data-set II. We note that the same set of coefficients both produce good agreement with the measurement.}
    \label{fig:bad_nps}
\end{figure*}

\figref{fig:base_nps_DS3522_487} and \figref{fig:base_nps_DS3564_487} show the total noise
simulation
along with the measured noise power spectrum (NPS) for both data-sets, before the amplifier gain but with the Bessel filter.
The system noise floor is set by the digitizer ADC range and thus changes with
different system gains for the two data-sets.
To give a better sense of the noise level, we also include the NPS plot of channel 474 in \figref{fig:base_nps_DS3564_474}, which is more representative of the CUORE channels.
In \tabref{tab:nps_sample_channel}, we have also listed the bias circuit settings and digitization system (i.e. the front-end board) data, such as the resistance of
the NTD-Ge thermistor at the biasing point, the system gain, and total
root-mean-square (RMS) voltage when there are no pulses for the sample channel.
The sample channel is selected as the same one in \secref{sec:dyn_model}.
We can observe that the Johnson noise from the bias resistor is the major
noise source.
However, the considered total noise only account for
approximately 1/10 of the measured NPS.
The figures also show that the continuous power spectrum contains a
$1/f$ component. This component cannot arise from the
white Johnson noise alone because
with the capacitance matrix in \eqnref{eqn:lin_sys_freq}, white noise input
spectrum always outputs $1/f^2$ power spectrum. We thus
add two extra input current noise terms:
\begin{equation}
    \label{eqn:noise}
    i^{2}_{\rm{ext}} = C_1 f+ \frac{C_2}{f}.
\end{equation}
\begin{table*}[!htb]
    \centering
    \caption{Bias circuit settings and digitization system data of channel 487 used in the NPS plots. The $V$ column lists the total biasing voltage of the electrical circuit in \figref{fig:electrical_circuit}. The gain column denotes the gain of the front-end amlification stage. The scaled baseline RMS is the raw baseline RMS scaled by the system gain. }
    \label{tab:nps_sample_channel}
    \adjustbox{width=\textwidth, center=\textwidth}{
    \begin{tabular}{c|CCCccc}
        \noalign{\vskip 1mm}
        \hline
        \hline
        \noalign{\vskip 1mm}
                    & \RL (R_{\rm{L}}) \ [\si{\giga \ohm}] & V \ [\si{\volt}] & \Rb \ [\si{\mega \ohm}] & Raw Baseline RMS [$\si{\micro \volt}$] & Gain & Scaled Baseline RMS [$\si{\milli \volt}$] \\[0.5ex]
        \hline
        \noalign{\vskip 1mm}
        Data-set I  & 60                                   & 4.71             & 250.3                   & 1.5                                    & 5150 & 7.9                                       \\
        Data-set II & 60                                   & 1.79             & 906.2                   & 1.3                                    & 2060 & 2.8                                       \\
        \hline
        \hline
    \end{tabular}
    }
\end{table*}
As shown in \figref{fig:bad_nps}, if we were to match the noise with the
measured NPS, the coefficient values are around $C_1 = \SI{7.1e-31}{\ampere^2 /
    \hertz^2 }$ and $C_2 = \SI{1.4e-30}{\ampere^2}$. These values are specific for
this channel, but we have taken observations on other fitted channels and they
behave in a similar way: the $C_1$ and $C_2$ values are around the same order
of magnitude as channel 487. We also observe that the same set of parameters
for one channel would produce good agreement with the measurements from both
data-sets.

\begin{figure}[!htb]
    \centering
    \includegraphics[width = 0.43\textwidth]{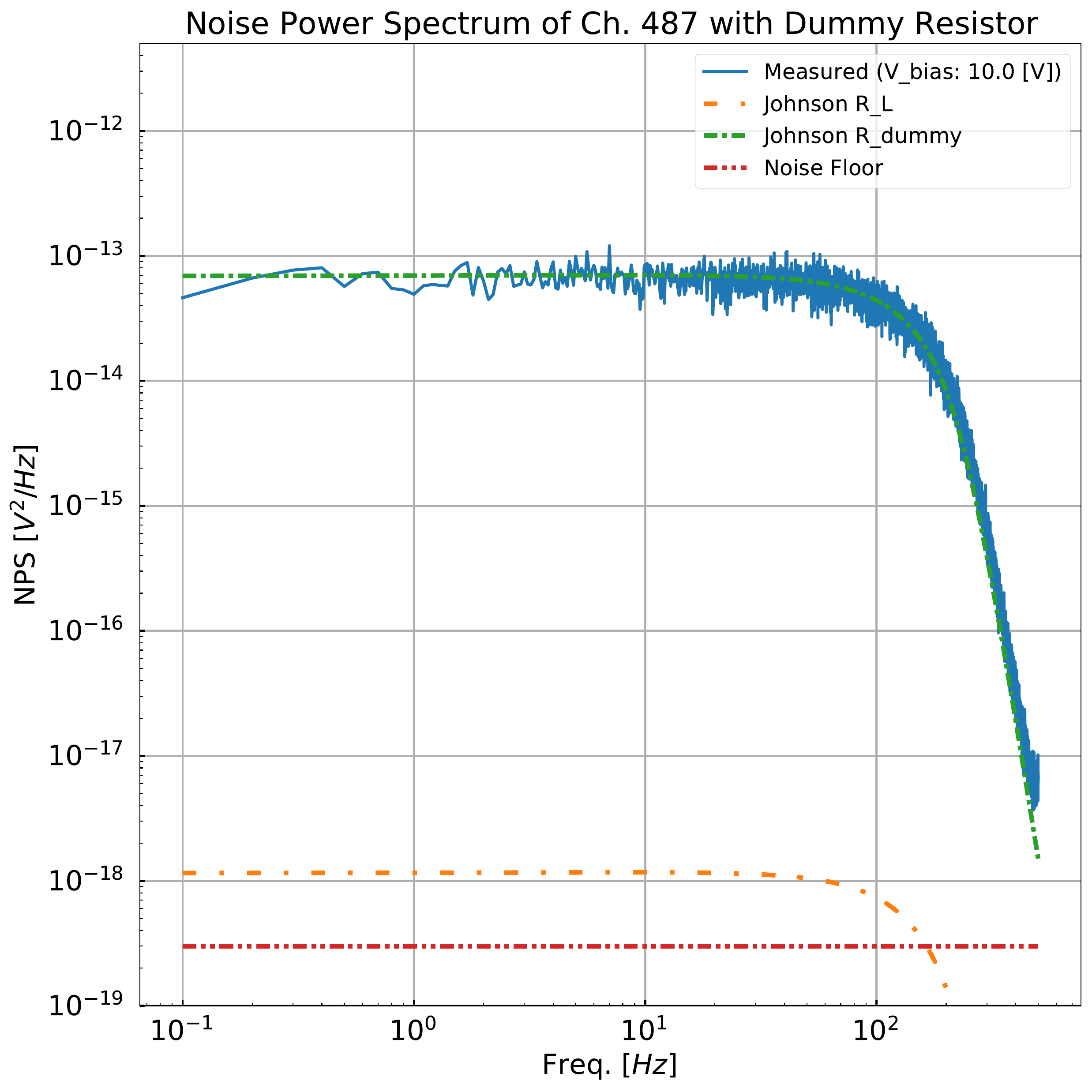}
    \caption{The simulated and measured NPS for channel 487's $\SI[]{4}[]{\mega \ohm}$ dummy resistor, along with $\RL = \SI{240}{\giga \ohm}$ (labelled $R_{\rm{L}}$) and a total biasing voltage of $\SI[]{10}[]{\volt}$. The white noise behavior with the Bessel filter roll-off agrees well with the data. }
    \label{fig:dummy_nps}
\end{figure}

Others \cite{amaboldi2002} have shown similar additional $1/f$ and $f$ noise
components that have to be considered in the input spectrum in large resistors:
\begin{equation}
    e^{2}_{J-L} =4kT\RL +\underbrace{K_{f}}_{\RL^2 C_1} f+ \underbrace{\beta V^{2}_{\rm{bias}} \RL}_{\RL^2 C_2} \frac{1}{f}.
\end{equation}
Specifically, the $\RL^2 C_2$ term is voltage and resistance
dependent.
The bias resistors of CUORE are custom designed to minimize noise and for them, $K_{f}$
and $\beta$ parameters are \cite{amaboldi2002}:
$K_{f} = \SI{5e-13}{\volt^2 / \hertz^2 }$ and
$\beta = \SI{1.96e-11}{\ohm^{-1}}$. We have found little change in the noise power
spectrum with these values. Moreover, we changed the bias resistors to higher
values (from $\SI{60}{\giga \ohm}$ to $\SI{240}{\giga \ohm}$)
in a few channels and saw no appreciable change in the NPS.
We have also measured the noise on channel 487's dummy resistor at room temperature
substituting the NTD-Ge. The dummy resistor is $\SI{4}{\mega \ohm}$ and is
under similar bias conditions of the NTD-Ges. The result showed only white
noise behavior for the dummy resistor, as shown in \figref{fig:dummy_nps}.
These tests suggest that the frequency-dependent noise sources mentioned in
\eqnref{eqn:noise} do not have resistance dependence.

\begin{figure}[!htb]
    \centering
    \includegraphics[width = 0.50\textwidth]{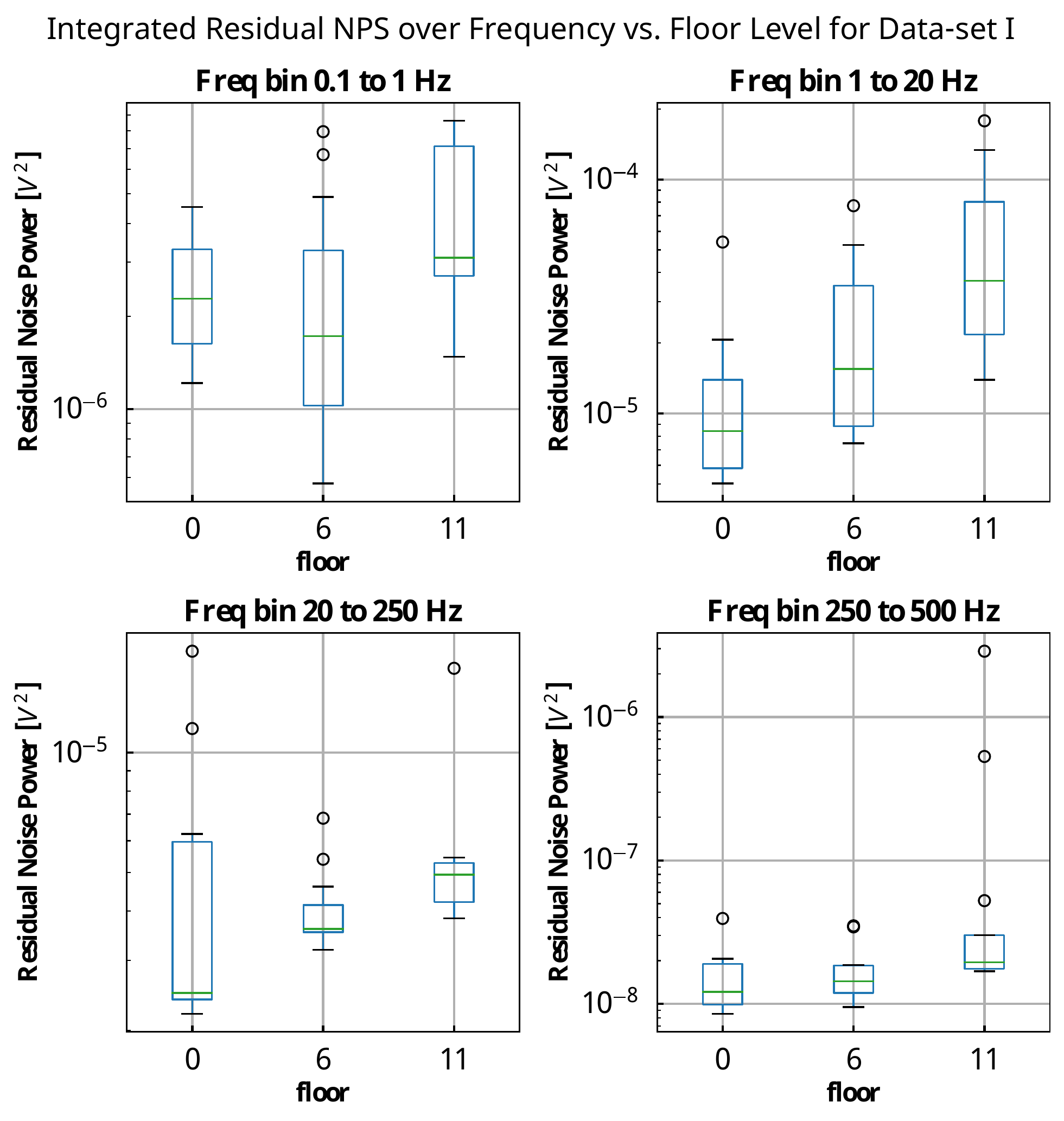}
    \caption{Residual noise power for the sample channels versus detector floor level. Higher floor level corresponds to positions closer to the mixing chamber plate. We note the trend that the closer the channels are to the mixing chamber plate, the more noisy they tend to become. The dots represent outlier channels with excessive integrated noise power.}
    \label{fig:resid_nps}
\end{figure}

We should mention that a hypothetical source of excess noise could be vibration
heating, as studied in another manuscript \cite{PIRRO2000331}. Although the
cryogenic set-up of CUORE is very different from the above work, vibrations are
known sources of excess noise for low temperature detectors
\cite{daddabbo2018a}. We integrated the residual noise power in the following
frequency bands: 0 to 1 Hz, 1 to 20 Hz, 20 to 250 Hz (excluding 50 Hz and
corresponding harmonics) and 250 to 500 Hz. The residual power for sample
channels of data-set I is plotted against the floor level in
\figref{fig:resid_nps}.
As the plot shows, the residual power increases with floor level in the
detector, which increases when the crystal is closer to cryostat thermal
stages.
Similar trend is also observed for data-set II.
The residual noise in the 20 to 250 Hz
band have almost no floor level dependence, suggesting that
other noise sources besides the vibrations may have contributed in the band.

It is difficult to ascertain $1/f$ component of the vibration noise from
flicker noise ($1/f^{-\gamma}$) because they are both non-stationary and
extremely correlated in nature. A detailed analysis of $1/f^{-\gamma}$ noise in
our experiment is out of scope for this paper and will be covered in future
studies.

\section{Conclusion}\label{sec:conclusion}

We have studied both the equilibrium and dynamic states of a
macroscopic calorimeter, and have analyzed our model's consistency
from both perspectives. We have confirmed that our three-node thermal model
is able to describe the equilibrium state of the detector, and that
for the NTD-Ge thermistors used by the CUORE experiment the thermistor
resistivity is a function of both the electron temperature and applied bias
voltage. We have found that the thermal coupling between the electron gas
and the Ge lattice is weaker than expected and dominates the electro-thermal
response of the calorimeter.

The dynamic model is able to predict the energy dependent pulse shape up to
at least $\SI{5407}{\kilo \electronvolt}$ with an additional
second-order temperature dependence correction term denoted as
the $q$ factor. This correction is approximately a few
percent when compared to the first order terms. We have made
several trials with different models, including ones with more
nodes and thermal couplings, and come to the conclusion that
the correction term is most likely the effect of a slight change
in the NTD-Ge resistivity law, for example an increase of the
$\gamma$ factor in \eqnref{eqn:ohmic_vrh} from the standard value
of 0.5. The fit
parameters for each channel are close even at two different
heat-sink temperatures, indicating that the dynamic model is consistent with
the equilibrium model.

As an application of the physical parameters obtained by the model, we
have also studied the noise power spectrum for the macro-calorimeters.
In the noise analysis, we find excess noise
power present in the measurements than expected, and modelled it.
Further study is necessary to determine the origin.
We, now, have verified that the residual noise power
increases with proximity to
the mixing chamber and the pulse tubes. Further usage of this model
could include the generation of simulated pulses for CUORE to implement
a machine-learning pulse energy identifier, and the potential to
optimize the detector response in the
CUORE Upgrade with Particle ID (CUPID) experiment.



\acknowledgments{
The CUORE Collaboration thanks the directors and staff of the Laboratori Nazionali del Gran Sasso and the technical staff of our laboratories.
This work was supported by the Istituto Nazionale di Fisica Nucleare (INFN); the National Science Foundation under Grant Nos. NSF-PHY-0605119, NSF-PHY-0500337,
NSF-PHY-0855314, NSF-PHY-0902171, NSF-PHY-0969852, NSF-PHY-1614611, NSF-PHY-1307204, NSF-PHY-1314881, NSF-PHY-1401832, and NSF-PHY-1913374; and Yale University.
This material is also based upon work supported by the US Department of Energy (DOE) Office of Science under Contract Nos. DE-AC02-05CH11231 and DE-AC52-07NA27344;
by the DOE Office of Science, Office of Nuclear Physics under Contract Nos. DE-FG02-08ER41551, DE-FG03-00ER41138, DE- SC0012654, DE-SC0020423, DE-SC0019316;
and by the EU Horizon2020 research and innovation program under the Marie Sklodowska-Curie Grant Agreement No. 754496.
This research used resources of the National Energy Research Scientific Computing Center (NERSC).
This work makes use of both the DIANA data analysis and APOLLO data acquisition software packages, which were developed by the CUORICINO, CUORE, LUCIFER and CUPID-0 Collaborations.

}

\section*{Data Availability}
Raw data were generated at the National Energy Research Scientific Computing
Center (NERSC) large scale facility.
Derived data supporting the findings of this study are available from the
corresponding author upon reasonable request.


\clearpage
\bibliographystyle{JHEP}
\bibliography{CUORE}

\end{document}